\documentclass[aps,prd,floats,floatfix,nofootinbib]{revtex4-1}
\usepackage{graphicx,amsmath}
\usepackage{amssymb}
\usepackage{amsfonts}
\usepackage{hyperref}
\usepackage[bottom]{footmisc}

\begin{document}

\title{Dilaton constraints and LHC prospects}

\author{Baradhwaj Coleppa}
\email {barath@physics.carleton.ca }

\author{Thomas Gr\'egoire}
\email {gregoire@physics.carleton.ca  }

\author{Heather E.\ Logan}
\email {logan@physics.carleton.ca }

\affiliation {Ottawa-Carleton Institute for Physics,
Carleton University,
Ottawa, Ontario K1S 5B6, Canada}

\date{November 14, 2011}

\begin{abstract}
The Standard Model Higgs searches using the first 1--2~fb$^{-1}$ of LHC data can be used to put interesting constraints on new scalar particles other than the Higgs.
We investigate one such scenario in which electroweak symmetry is broken via strongly coupled conformal dynamics.  This scenario contains a neutral scalar dilaton---the Goldstone boson associated with spontaneously broken scale invariance---with a mass below the conformal symmetry breaking scale and couplings to Standard Model particles similar (but not identical) to those of the Standard Model Higgs boson.  We translate the LEP and LHC Higgs limits to constrain the dilaton mass and conformal breaking scale.  The conformal breaking scale $f$ is constrained to be above 1~TeV for dilaton masses between 145 and 600~GeV, though it can be as low as 400~GeV for dilaton masses below 110~GeV.
We also show that (i) a dilaton $\chi$ with mass below 110~GeV and consistent with the LEP constraints can appear in $gg \to \chi \to \gamma\gamma$ with a rate up to $\sim$10 times the corresponding Standard Model Higgs rate, and (ii) a dilaton with mass of several hundred GeV is much narrower than the corresponding Standard Model Higgs, leading to improved search sensitivity in $\chi \to ZZ \to 4\ell$.
\end{abstract}

\maketitle
\section{Introduction}

The main goal of the CERN Large Hadron Collider (LHC) is to uncover the agent of electroweak symmetry breaking (EWSB). Over the years many possibilities have been envisioned which generally fall into one of two categories: weakly coupled physics, or strongly coupled physics. Weakly coupled EWSB typically requires a Higgs doublet that acquires a vacuum expectation value (vev), the prototypical example of which is the Standard Model (SM) that has a light scalar Higgs boson with couplings to other particles proportional to their masses.  Strongly coupled EWSB on the other hand does not require a light scalar, but in general predicts a large number of bound states above the electroweak scale. Electroweak precision tests tend to favor a weakly coupled light Higgs; however, this scenario suffers from severe fine-tuning. To alleviate the fine-tuning problem the light Higgs must be part of a larger sector, such as in supersymmetric models \cite{Dimopoulos:1981zb} or little Higgs models~\cite{LH}. Most such models still require some amount of fine-tuning to evade experimental bounds. Fortunately the two classes of scenarios have qualitatively different spectra, and the LHC should be able to easily distinguish between them. 

If the strongly coupled theory is conformal, however, the low energy theory might include a light scalar dilaton, which has properties that are similar to those of a light Higgs~\cite{Gildener:1976ih,Goldberger:2007zk}. Conformality of the strongly interacting sector that breaks the electroweak symmetry~\cite{walking} has various motivations. It helps, for example, to avoid problems with flavor-changing neutral currents that potentially plague theories of extended technicolor~\cite{etc}. It can also help the agreement of these kinds of theories with electroweak precision tests~\cite{etcS}. The properties of the ``techni-dilaton'' appearing in theories of this kind and the associated LHC phenomenology were studied in Ref.~\cite{technidilaton}.  Similar phenomenology was also studied in the context of Randall-Sundrum (RS)~\cite{hep-ph/9905221}  models as the AdS-CFT correspondence~\cite{hep-th/9711200} dictates that the radion in RS scenarios is dual to a dilaton of the associated CFT. In fact, the couplings of the radion to the SM sector are the same as those of the dilaton, except for a contribution due to the interaction of the radion with the ``bulk'' of the extra dimension that can contribute significantly to the radion couplings to gluon or photon pairs. From the CFT point of view, this contribution arises because the SM gauge bosons are not part of the CFT but instead come from weakly gauging some of the global symmetries of the CFT. Reference~\cite{arXiv:0705.3844} discusses the physics of the radion in detail, while also establishing the equivalence of the CFT and RS pictures.   Constraints on the radion scenario from recent LHC Higgs searches have been studied very recently in Ref.~\cite{radionLHC}.

In this paper, we consider a scenario in which the scale invariance of the strong dynamics is manifest at very high energy but is spontaneously broken at a scale $f$, not too far above the electroweak scale. The strong dynamics is also responsible for the breaking of  the electroweak symmetry at scale $v = 246$~GeV. We assume that the sector that explicitly breaks the conformal symmetry has a small parameter which makes the dilaton---the Goldstone boson associated with the spontaneously broken scale symmetry---parametrically lighter than the other resonances, which are expected to have masses around $4 \pi f$. Finally we imagine that the entire structure of the SM is embedded in the conformal sector at high energy. This is the scenario presented in Refs.~\cite{Goldberger:2007zk, Fan:2008jk}, and for which the dilation couplings can be deduced from symmetry arguments. 

We do not consider the effects of Higgs-dilaton mixing, or the decay of the dilaton to two Higgses, the reason being that we would like to study a scenario in which the electroweak symmetry is broken by strong dynamics and not by a weakly coupled Higgs. Note however that we will consider rather high values for $f$ compared to the SM electroweak breaking scale $v = 246$~GeV. In this case there might be an additional Higgs-like state with mass smaller than $f$, like for example in the strongly-coupled light Higgs scenario of Ref.~\cite{Giudice:2007fh,Vecchi:2010gj}. We will assume that such a state is heavier than the dilaton and does not mix significantly with it. 

The paper is organized as follows.  We start by defining the dilaton couplings and deriving the dilaton decay width and branching ratios in Sec.~\ref{sec:couplings}.  We  determine the existing constraints on the dilaton mass and the conformal breaking scale $f$ from SM Higgs searches at the CERN Large Electron-Positron Collider (LEP) and the LHC in Sec.~\ref{sec:constraints}.  
We then propose two new LHC analyses that could increase the sensitivity to a dilaton in Sec.~\ref{sec:discovery}.  We also discuss how a dilaton can be distinguished from the SM Higgs based on rates in different detection channels.
We finish with some brief comments on dilaton production at a future International Linear $e^+e^-$ Collider (ILC) and its photon collider variant in Sec.~\ref{sec:ilc} and conclude in Sec.~\ref{sec:conclusions}.

\section{Properties of the dilaton}
\label{sec:couplings}

\subsection{Dilaton couplings}
The dilaton is the Goldstone boson associated with spontaneously broken scale invariance and can be introduced in the low energy Lagrangian as a compensator for scale transformations---i.e., all couplings that are not scale invariant can be made scale invariant by introducing appropriate powers of the dilaton field to compensate for the shift under a scale transformation.  Following this prescription, at energies below the conformal breaking scale, the coupling of the dilaton field $\chi$ to the SM sector is described by the Lagrangian~\cite{Goldberger:2007zk,Fan:2008jk}\footnote{Ref. \cite{Fan:2008jk} discusses ``anomalous'' couplings of the dilaton to fermion pairs that are not proportional to the fermion masses---we ignore this possibility here.} 
\begin{equation}
	\mathcal{L}=\frac{v^2}{4} \textrm{Tr} \left|D_{\mu}U\right|^2 \left(\chi/f\right)^2-\frac{1}{4}\left(B_{\mu\nu}\right)^2-\frac{1}{2}\textrm{Tr}\left(W_{\mu\nu}\right)^2 -m_{i}\bar{\psi}_{i}U\psi_{i}\left(\chi/f\right),
	\label{eq:invariant electroweak chiral}
\end{equation}
where $U$ is a $2\times 2$ non-linear signal model field  given by $U=\textrm{exp}[i (\pi^a \tau^a /v) (f/\chi)]$, where $\pi^a$ are the electroweak Goldstone bosons eaten by the $W$ and $Z$, and $\tau^a$ are the SU(2) generators. The first term contains the mass term of the gauge bosons and the coupling of the gauge bosons to the dilaton. To compute this coupling, let us parametrize the fluctuations of the physical dilaton $\bar \chi$ about its vev as $\bar{\chi}=\chi-f$.  
We can now directly read off the couplings of the dilaton to the gauge bosons and fermions from Eq.~(\ref{eq:invariant electroweak chiral}).  The dilaton couplings of dimension up to four are given by,
\begin{equation}
	\mathcal{L}_{\chi,SM}=\frac{1}{2}M_{V}^2 V_{\mu}^2\left(\frac{2\bar{\chi}}{f}+\frac{\bar{\chi}^2}{f^2}\right)-\frac{\bar{\chi}}{f} m_{i}\bar{\psi}_{i}\psi_{i}.
	\label{eq:dilaton couplings}
\end{equation}
The three-point couplings are identical to those of the Higgs boson in the SM multiplied by an overall $v/f$ scaling factor.

The dilaton couples to any term of the Lagrangian that breaks scale invariance. More formally, it couples ${T_\mu}^\mu$, the trace of the stress-energy tensor. This includes the scale anomaly, proportional to the $\beta$ function of the Standard Model.\footnote{More precisely, the divergence of the dilatation current is related to the trace of the energy-momentum tensor $T^{\mu}_{\mu}$, which  can be computed by evaluating the trace anomaly.  In QCD this gives $T^{\mu}_{\mu} = (\beta_G/g_s^3)(G_{\mu\nu}^a)^2$, while in QED it gives $T^{\mu}_{\mu} = (\beta_{EM}/2e^3)(F_{\mu\nu}^2)$.} Indeed, within the SM, the running of the QCD and QED gauge couplings introduces a dependence on a renormalization scale and this breaks scale invariance at loop level; this effect is proportional to the appropriate beta functions.  This induces the direct couplings $\chi G_{\mu\nu}^2$ and $\chi F_{\mu\nu}^2$ that are given by
\begin{equation}
	\mathcal{L}=\left[-\frac{\alpha_{EM}}{8\pi}b_{EM}(F_{\mu\nu})^2-\frac{\alpha_{s}}{8\pi}b_{G}(G_{\mu\nu}^a)^2 \right]\textrm{ln}\frac{\chi}{f},
\end{equation}
where the coefficients $b_{EM}$ and $b_{G}$ are the beta function coefficients to be evaluated at the energy scale of the interaction.  For an on-shell physical dilaton, this corresponds to including all particles lighter than the dilaton in the beta function coefficient.  

A second, less mysterious way to understand these couplings is to consider the beta functions of the SM gauge interactions above the scale of conformal symmetry breaking.  If the SM gauge interactions are part of the conformal sector, as we assume here, their gauge couplings must not run above the conformal symmetry breaking scale; i.e., $b_G = b_{EM} = 0$ at the high scale.  This is achieved through the presence of new, gauge-charged states that cancel the SM contribution to the beta functions and get masses through conformal breaking around $4 \pi f$; in particular,
\begin{equation}
	\sum_{\rm light} b_i + \sum_{\rm heavy} b_i = 0,
	\label{eq:lightheavy}
\end{equation}
where the sums run over the contributions of the SM fields and the conformal sector fields, respectively.  From this it becomes clear how to compute the dilaton coupling to two gluons or two photons from a loop-calculation perspective.  The heavy particles above the conformal-breaking scale run in the loops and give a contribution to the coupling proportional to $\sum_{\rm heavy} b_i$, which can be rewritten in terms of the SM beta function coefficients using Eq.~(\ref{eq:lightheavy}).  The usual SM particles also run in the loops and their contributions have to be included in the computation of branching ratios and cross-section as we will show later.  We thus obtain the effective Lagrangian,
\begin{equation}
	 \mathcal{L} = \left[ \frac{\alpha_{EM}}{8\pi} 
	 \left( - b_{EM}
 \right)(F_{\mu\nu})^2
	 + \frac{\alpha_{s}}{8\pi} \left( - b_{G} 
\right)
	 	(G_{\mu\nu}^a)^2 \right]
		\frac{\bar \chi}{f},
	\label{eqn:massless couplings}
\end{equation}
where we have used Eq.~(\ref{eq:lightheavy}) to swap the heavy-particle beta function coefficients for the familiar SM beta function coefficients (including the top quark), $b_{EM} = -11/3$ and $b_G = 11 - \frac{2}{3} n_f$, with $n_f = 6$. Note the $1/f$ dependence of the dilaton coupling in Eq.~(\ref{eqn:massless couplings}) in place of the $1/v$ dependence of the corresponding SM Higgs coupling. We also note that the argument given above makes it clear that in models where the SM gauge groups are not part of the conformal dynamics, but are instead a weak perturbation of it, the coupling of the dilaton to gauge bosons will be different.

\subsection{Cross sections, decay widths, and branching ratios}

\begin{figure}
\includegraphics[scale=0.3]{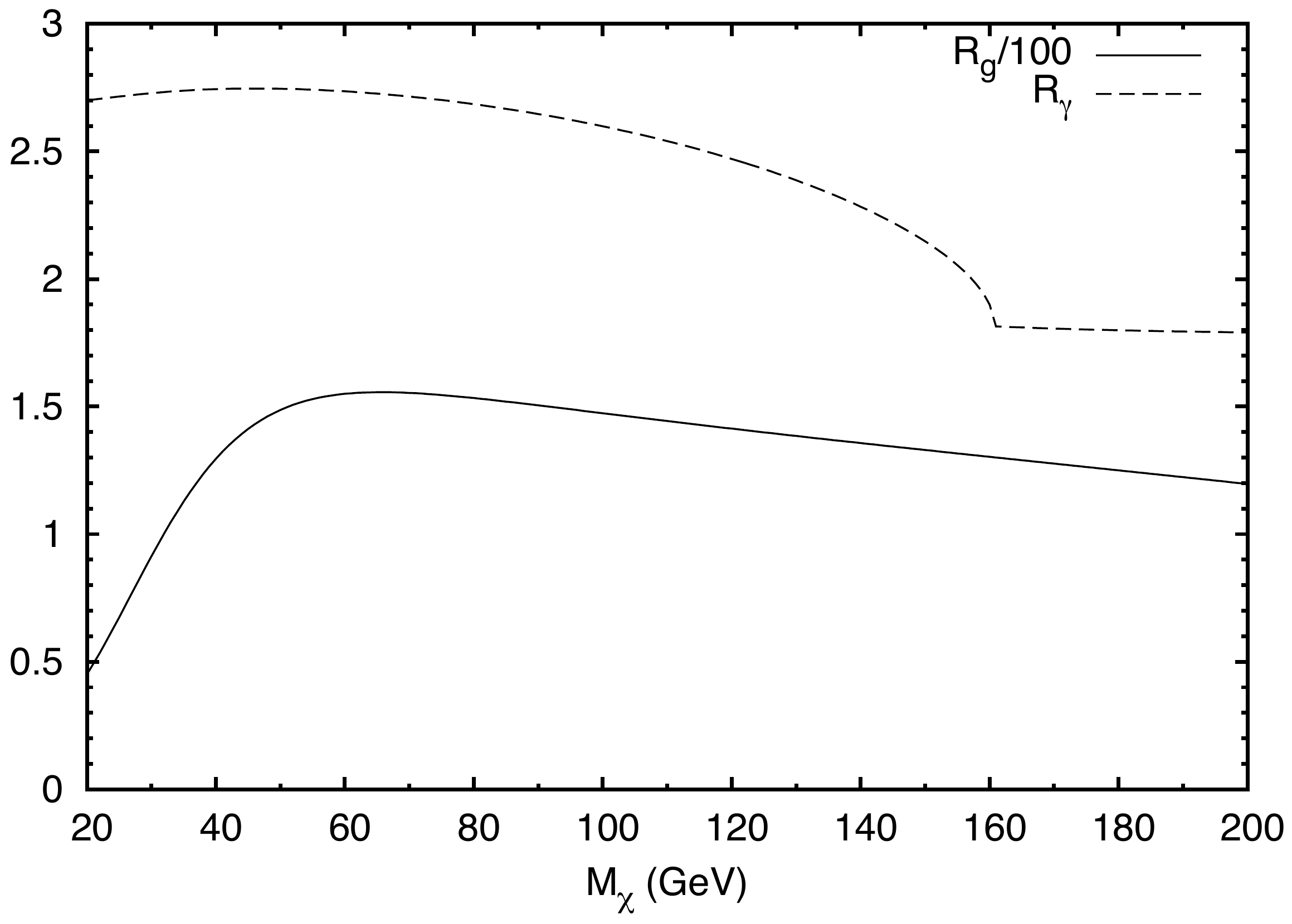}
\includegraphics[scale=0.3]{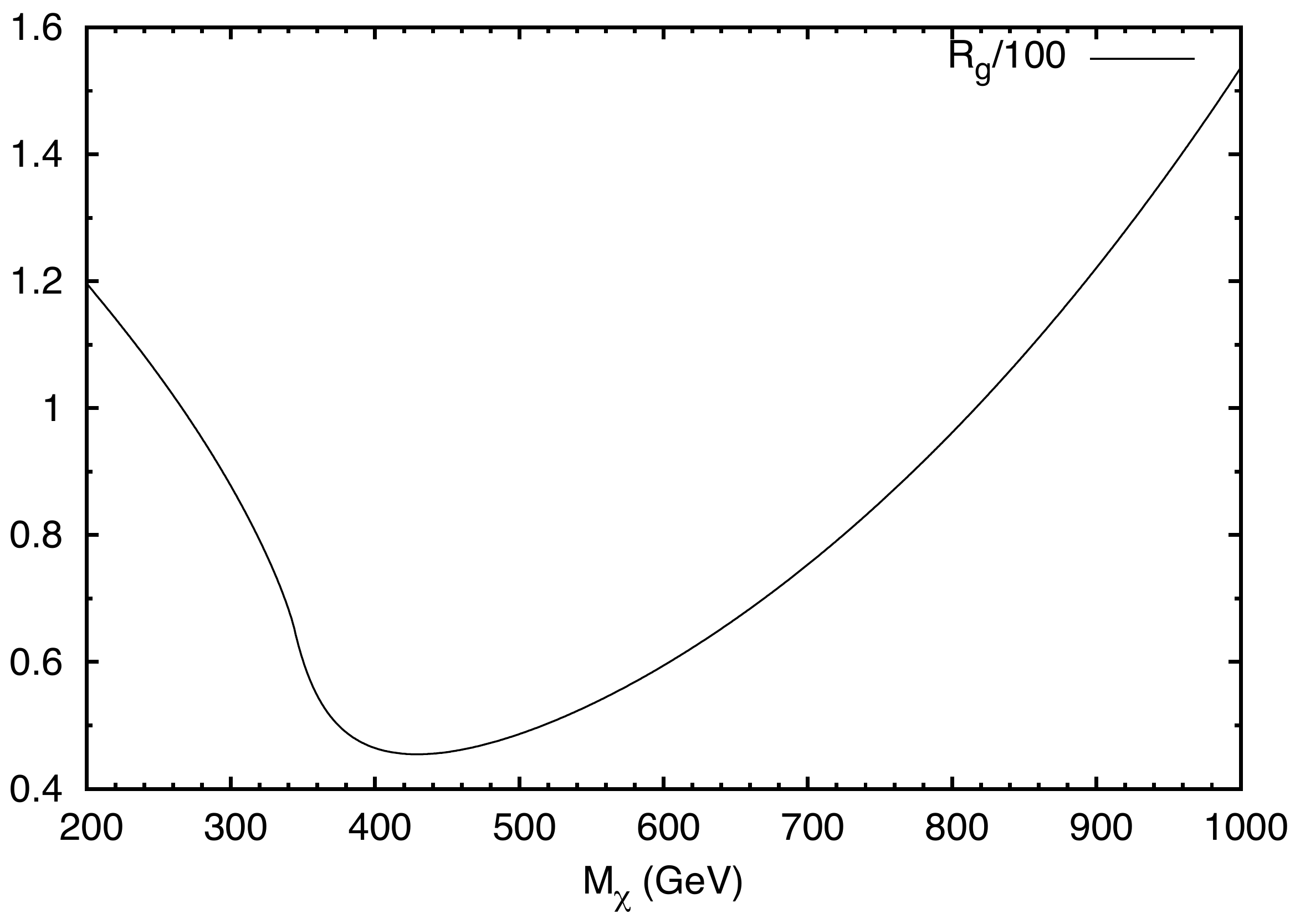}
\caption{The scaling functions $R_g$ and $R_{\gamma}$ as a function of the dilaton mass.  Note that $R_g$ is divided by 100 in the plots to allow the two functions to be displayed on the same axes.}
\label{fig:R}
\end{figure}

The behavior of the dilaton couplings compared to those of the SM Higgs is easy to understand.  The partial widths for the dilaton to decay into any massive SM final state, and the cross section for production of the dilaton from its coupling to massive SM particles, are all equal to their SM values times an overall scaling factor of $v^2/f^2$.  The partial width for the dilaton to decay to two gluons (or two photons) and the cross section for production of the dilaton via gluon fusion (or two-photon fusion), are equal to their SM values times an overall scaling factor of $R_g v^2/f^2$ (or $R_{\gamma} v^2/f^2$), where the function $R_g$ (or $R_{\gamma}$) is the ratio of the gluon (or photon) loop factor squared for the dilaton to that of the SM Higgs:
\begin{eqnarray}
	R_g &=& \frac{\left| -b_G + \frac{1}{2} \sum_i F_{1/2}(\tau_i) \right|^2}
	{\left| \frac{1}{2} \sum_i F_{1/2}(\tau_i) \right|^2}, \nonumber \\
	R_{\gamma} &=& \frac{\left| -b_{EM} + \sum_i N_{ci} Q_i^2 F_i(\tau_i) \right|^2}
	{\left| \sum_i N_{ci} Q_i^2 F_i(\tau_i) \right|^2}, 
	\label{eq:R}
\end{eqnarray}
where $N_{ci}$ is the number of colors and $Q_i$ is the electric charge in units of $e$ for particle $i$ running in the loop.  The loop functions $F_1(\tau)$ and $F_{1/2}(\tau)$, for vector bosons and fermions respectively running in the loop, are given by~\cite{Gunion:1989we}:
\begin{eqnarray}
	F_{1}&=&2+3\tau+3\tau(2-\tau)f(\tau) \nonumber \\
	F_{1/2}&=&-2\tau\left[1+(1-\tau)f(\tau) \right],
	\label{eq:F}
\end{eqnarray}
where $\tau_i=4 m_i^2/M_{\chi}^2$ and
\begin{equation}
f(\tau) =
  \begin{cases}
   \left[\sin^{-1}\left(\sqrt{1/\tau} \right) \right]^2 & \text{if } \tau \geq 1 \\
   -\frac{1}{4}\left[\textrm{ln}(\eta_+/\eta_-)-i\pi  \right]^2      & \text{if } \tau < 1,
  \end{cases}
\label{eqn:tau}
\end{equation}
with $\eta_{\pm}=(1\pm\sqrt{1-\tau})$.  The functions $R_g$ and $R_{\gamma}$ are plotted in Fig.~\ref{fig:R} as a function of the physical dilaton mass $M_{\chi}$.  (From here on we drop the bar on the physical dilaton and call it $\chi$ for simplicity.)  For dilaton masses above 100~GeV, only the top quark and $W$ boson loops have a significant numerical effect in Eq.~(\ref{eq:R}).  For masses below 100~GeV we include also the bottom, charm, and tau loops.\footnote{We use $M_W = 80.4$~GeV, $m_t = 172$~GeV, $m_b = 4.2$~GeV, $m_c = 1.4$~GeV, and $m_{\tau} = 1.777$~GeV when evaluating $R_g$ and $R_{\gamma}$.}  Note that $R_{\gamma}$ stays between 1.75 and 2.75 for $M_{\chi}$ between 20~GeV and 200~GeV, while $R_g$ varies between 40 and 160 for $M_{\chi}$ between 20~GeV and 1000~GeV.

\begin{figure}
\includegraphics[scale=0.31]{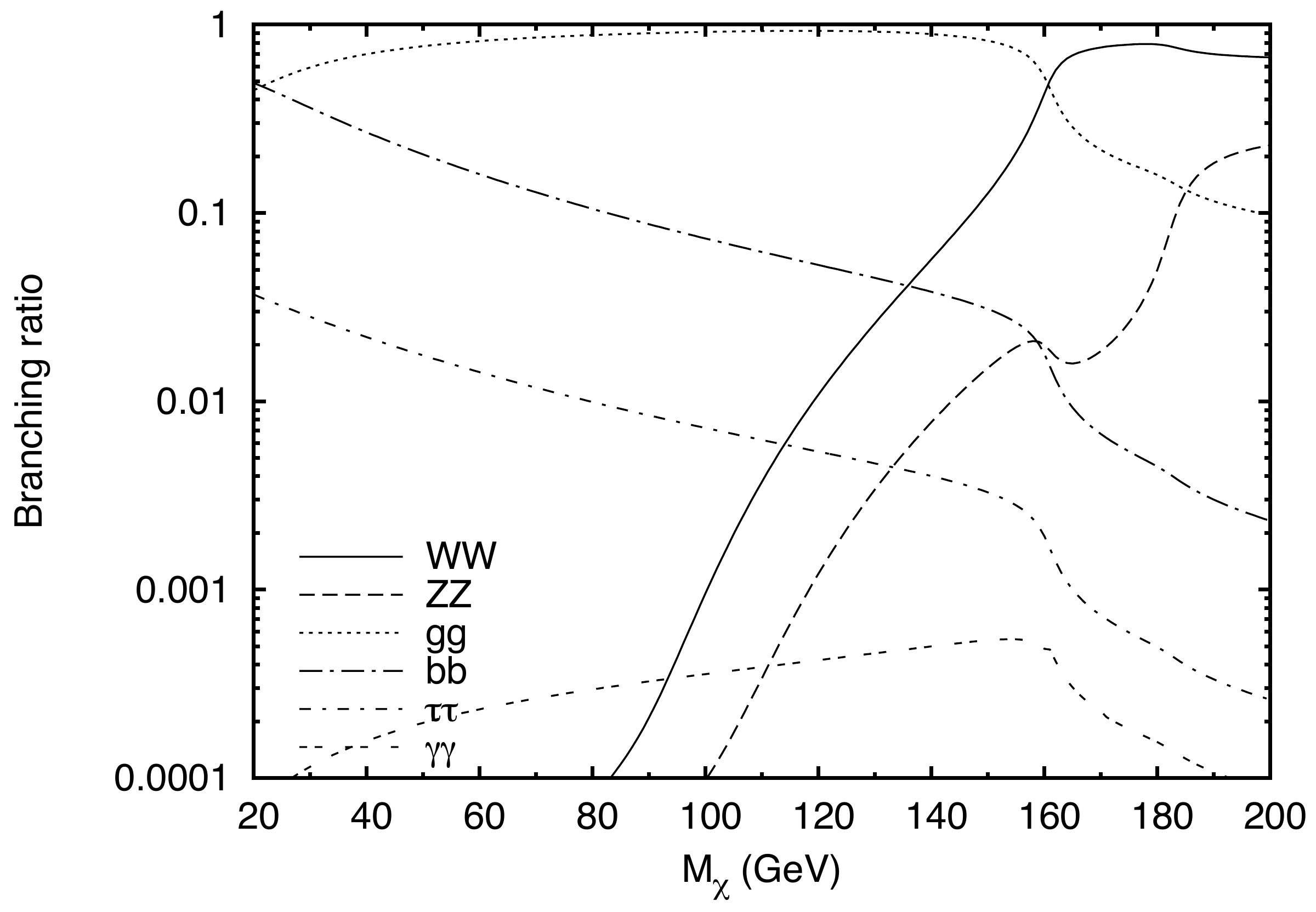}
\includegraphics[scale=0.31]{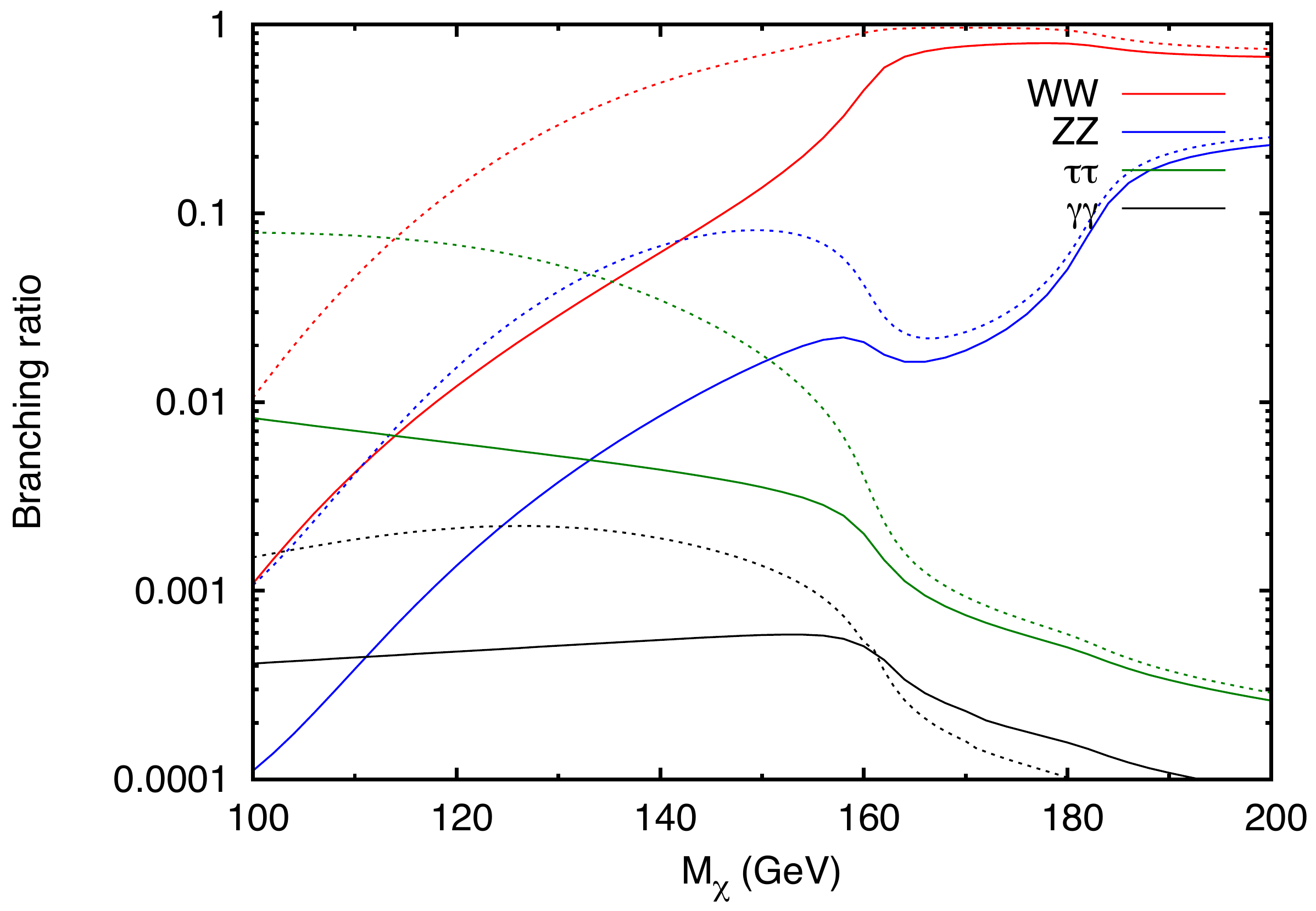}
\caption{Dilaton branching ratios as a function of the dilaton mass for masses below 200~GeV.
The right-hand plot compares the dilaton branching ratios into final states important for LHC Higgs searches in inclusive production modes (solid lines) to the corresponding SM Higgs branching ratios (dotted lines).}
\label{fig:dil-BRlight}
\end{figure}

\begin{figure}
\includegraphics[width=5in]{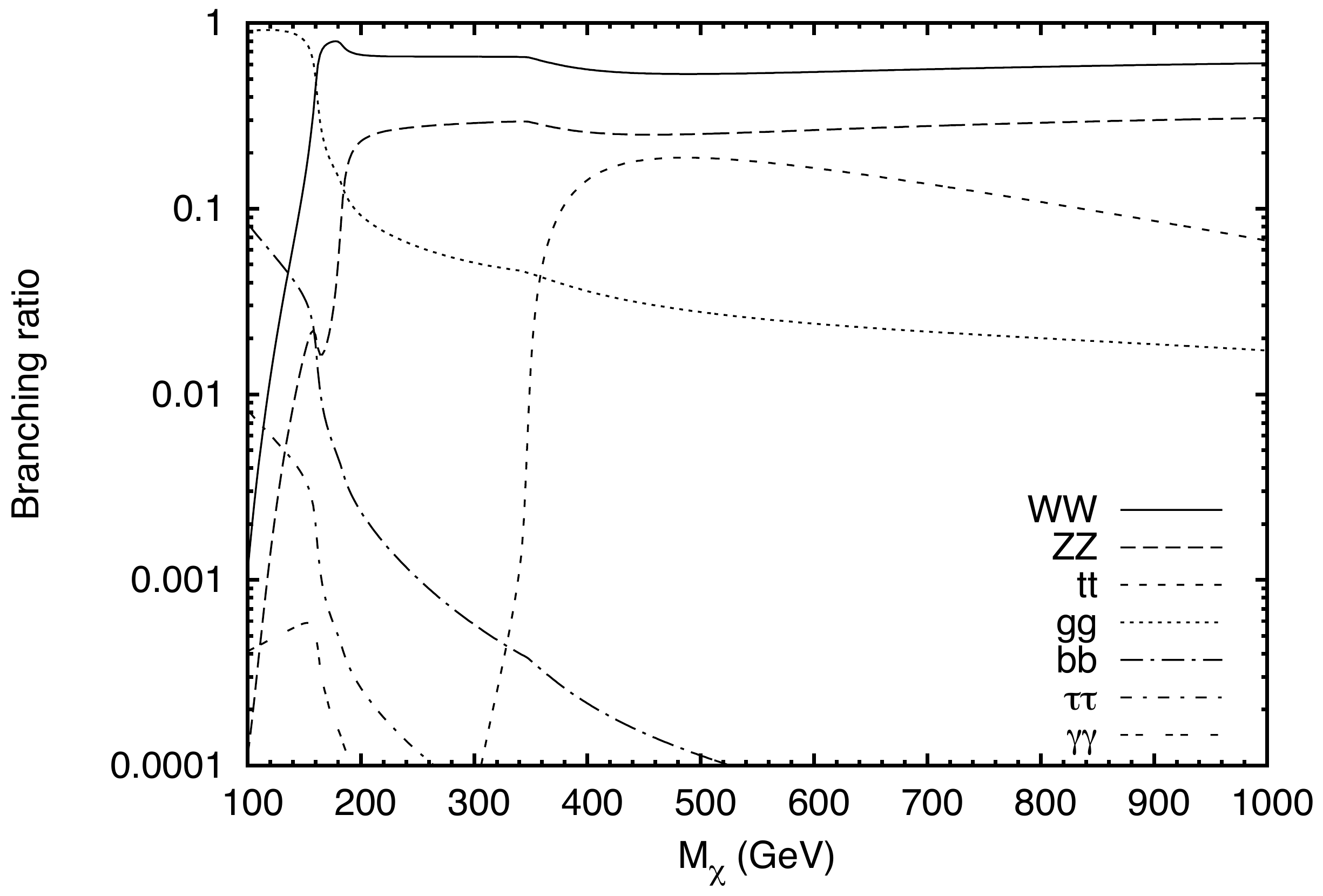}
\caption{Dilaton branching ratios as a function of the dilaton mass for masses up to 1000~GeV.}
\label{fig:dil-BRfull}
\end{figure}

\begin{figure}
\includegraphics[width=5in]{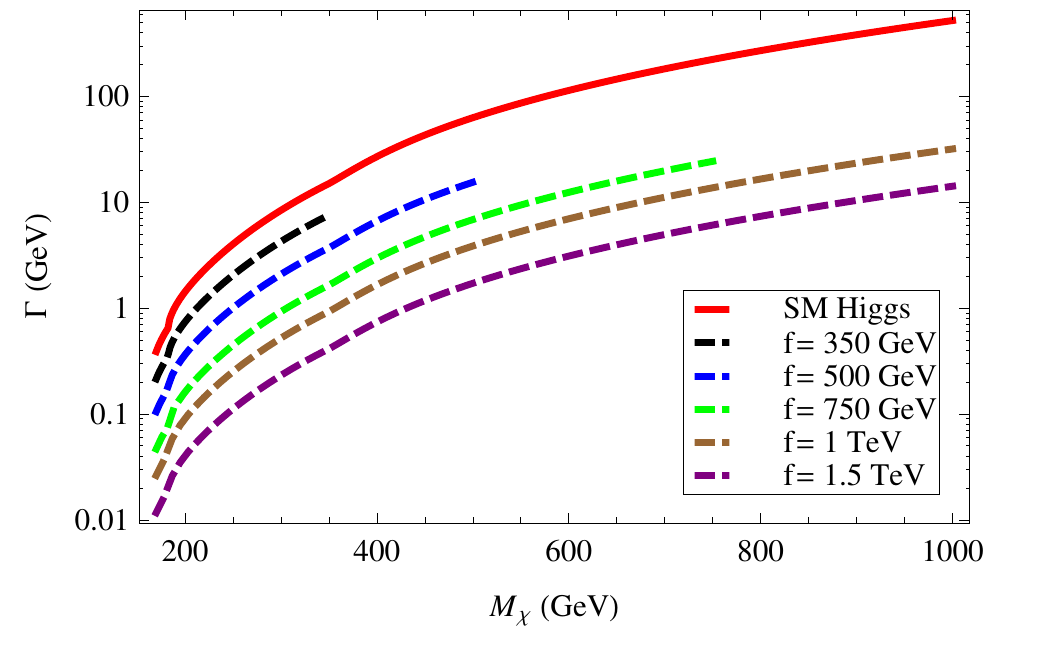}
\caption{Total decay width of the dilaton as a function of the dilaton mass, for various values of the conformal symmetry breaking scale $f$.  We plot only $M_{\chi} < f$.  The corresponding total width of the SM Higgs is shown for comparison (solid line).}
\label{fig:width}
\end{figure}

Because of the close correspondence between the dilaton couplings and those of the SM Higgs, we can compute the dilaton decay branching ratios and total width by rescaling the known SM Higgs partial widths to SM final states.\footnote{\label{fn:rcs}This approach allows us to incorporate the known radiative corrections to the SM Higgs partial widths.  These radiative corrections transfer over to the dilaton \emph{exactly} except for (i) electroweak radiative corrections that involve the triple-Higgs vertex or more than one coupling of a Higgs to other SM particles, and (ii) QCD and electroweak radiative corrections to the $Hgg$ and $H\gamma\gamma$ vertices that resolve the top and/or $W$ loops rather than treating the coupling as a pointlike effective vertex.}  
We compute the SM Higgs decay partial widths using {\tt HDECAY}~3.53~\cite{HDECAY}.  We find the dilaton partial widths at the corresponding mass point by rescaling all the SM Higgs partial widths by $v^2/f^2$, with an additional scaling by $R_g$ ($R_{\gamma}$), given in Eq.~(\ref{eq:R}), for decays to $gg$ ($\gamma\gamma$).  The resulting dilaton branching ratios are plotted in Figs.~\ref{fig:dil-BRlight} and \ref{fig:dil-BRfull} for dilaton masses below 200~GeV and up to 1000~GeV, respectively.  These branching ratios are independent of the conformal breaking scale $f$.  The total width of the dilaton, proportional to $v^2/f^2$, is plotted for various $f$ values in Fig.~\ref{fig:width}.

We observe the following features.  A dilaton with mass less than $2 M_W$ decays predominantly to a pair of gluons.  This is due to the dramatic enhancement of the $\chi gg$ coupling via the QCD beta function coefficient.  In particular, this suppresses the dilaton branching ratios to $b \bar b$, $\tau\tau$, and $\gamma\gamma$ below the $WW$ threshold compared to the corresponding branching ratios of the SM Higgs.  It also suppresses the branching ratio to off-shell $WW$ and $ZZ$ below $2 M_W$.  These features will have a significant impact on the LEP and LHC dilaton exclusions below $2 M_W$.  

For dilaton masses above $2 M_W$, the branching ratios become essentially identical to those of the SM Higgs, with the dilaton decaying predominantly to $WW$, $ZZ$, and $t \bar t$.  Decays to gluon pairs contribute at most 10\% at $M_{\chi} = 200$~GeV, falling to less than 2\% at $M_{\chi} = 1000$~GeV.  The partial widths to the dominant $WW$, $ZZ$, and $t \bar t$ decay modes are equal to the corresponding SM Higgs partial widths times a scaling factor $v^2/f^2$.  This leads to a dramatic suppression of the dilaton total width at large $f$ values compared to that of the SM Higgs.

\section{Constraints from Higgs searches}
\label{sec:constraints}

\subsection{Constraints from LEP}

The LEP experiments searched for Higgs production in $e^+e^- \to Z H$ via an intermediate off-shell $Z$ boson.  The SM Higgs limits~\cite{LEPfinal} are based on the SM Higgs decay final states $b \bar b$ and $\tau \tau$, which dominate in the SM Higgs mass range to which LEP had kinematic access.  LEP presented these results as an exclusion in the parameter space of $M_H$ and $\xi^2$, where $\xi$ is a scaling factor on the $ZZH$ production coupling, assuming SM decay branching ratios.  

\begin{figure}
\includegraphics[scale=0.5]{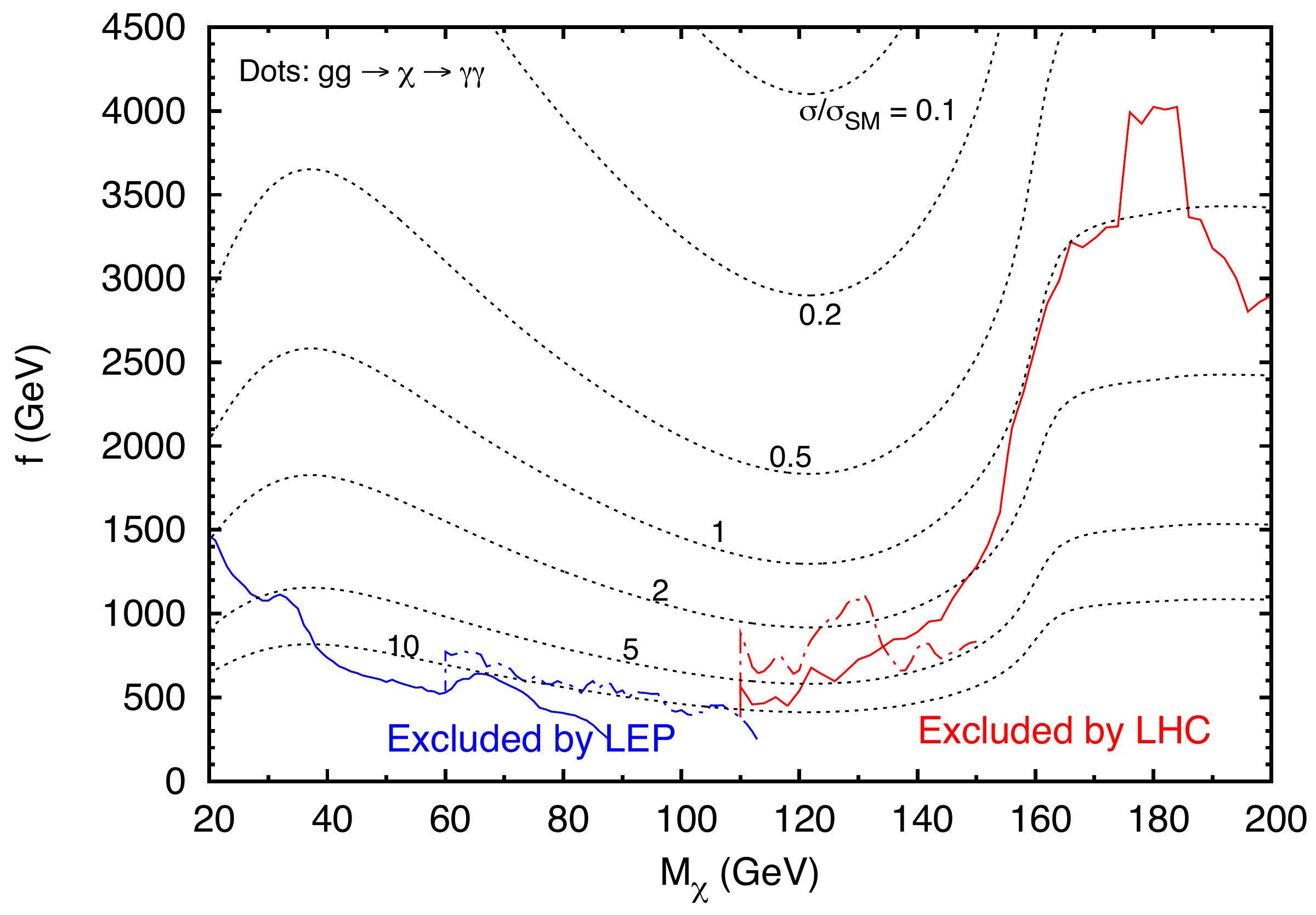}
\caption{Constraints on the dilaton mass and conformal scale $f$ from LEP and the LHC.  Regions below the solid and dot-dashed blue lines are excluded by the LEP SM Higgs search and a flavor-independent LEP search for a hadronically-decaying Higgs, respectively.  Regions below the solid and dot-dashed red lines are excluded by the combined LHC SM Higgs search and the LHC Higgs search in the $H \to \gamma\gamma$ channel, respectively.  For the LHC exclusion, we take the stronger of the ATLAS and CMS limits at each mass point.  The dotted black lines show contours of $\sigma(gg \to \chi \to \gamma\gamma)/\sigma(gg \to H_{\rm SM} \to \gamma\gamma)$.}
\label{fig:lep}
\end{figure}

We translate the LEP combined SM Higgs limit of Ref.~\cite{LEPfinal} into a limit on the dilaton by identifying the scaling factor $\xi^2$ as
\begin{equation}
	\xi^2 = \frac{\sigma(e^+e^- \to Z \chi)}{\sigma(e^+e^- \to Z H_{\rm SM})} \times
	\frac{{\rm BR}(\chi \to b \bar b + \tau\tau)}{{\rm BR}(H_{\rm SM} \to b \bar b + \tau\tau)}.
\end{equation}
The cross section ratio is equal to $v^2/f^2$, while we calculate the ratio of branching ratios as described in Sec.~\ref{sec:couplings}.  The double suppression---by $v^2/f^2$ on one hand and by the suppressed dilaton branching ratios into $b \bar b$ and $\tau\tau$ on the other---leads to a rather weak dilaton limit from the LEP SM Higgs search.  This is shown by the solid blue line in Fig.~\ref{fig:lep}.

The LEP experiments also performed a flavor-independent search for $e^+e^- \to ZH$ with the Higgs decaying into hadrons, for Higgs masses above 60~GeV~\cite{LEPhadronic}.  This search is more sensitive to dilaton production than the SM Higgs search because it captures the dominant branching ratio to gluons.  The limit was again presented as an exclusion in the parameter space of $M_H$ and a scaling factor $\xi^2$, defined for our purposes as
\begin{equation}
	\xi^2 = \frac{\sigma(e^+e^- \to Z \chi)}{\sigma(e^+e^- \to Z H_{\rm SM})} \times
	{\rm BR}(\chi \to gg + b \bar b + c \bar c).
\end{equation}
This search leads to the strongest limit on the dilaton in the mass range 60--110~GeV, shown by the dot-dashed blue line in Fig.~\ref{fig:lep}.  Together, these LEP limits exclude values of $f$ below 400~GeV.\footnote{A similar analysis of LEP constraints on the dilaton was done in Ref.~\cite{arXiv:0909.1319} using the public code {\tt HiggsBounds} 1.0~\cite{higgsbounds}, resulting in constraints similar to what we obtain from the LEP combined limit.  Ref.~\cite{arXiv:0909.1319} also determined the dilaton exclusion from the Tevatron $H \to WW$ search data; this is now superseded by the LHC results.}

\subsection{Constraints from the LHC}

The LHC experiments have placed strong exclusions on the mass of the SM Higgs by combining various channels, the most important of which are $\gamma\gamma$, $WW\rightarrow \ell\nu \ell\nu$, and $ZZ\rightarrow 4\ell,\ell\ell\nu\nu$. The ATLAS and CMS collaborations together now exclude a Higgs with SM couplings in the mass range 145--466 GeV, with a small window at 288--296~GeV which is disfavored but not quite excluded at 95\% confidence level~\cite{ATLAScomboLP11,CMScomboLP11}.  The dilaton appears in the same search channels as the SM Higgs, but with different production rates and decay branching ratios due to the modification of its couplings relative to those of the SM Higgs.  As we have seen, the dilaton coupling to gluon pairs can be dramatically enhanced.  This leads to enhanced dilaton production in gluon fusion for a large range of $f$ values, allowing us to set strong limits, especially above the $WW$ threshold where the dilaton branching ratios to final states used in the LHC searches are not suppressed. 

We translate the current LHC exclusion on the SM Higgs to an exclusion on the dilaton as follows. We compute the inclusive dilaton production cross section by scaling the SM Higgs cross sections from gluon fusion and vector boson fusion (VBF) according to
\begin{equation}
	\frac{\sigma(pp \to \chi)}{\sigma(pp \to H_{\rm SM})} 
	= \frac{\sigma(gg \to \chi) + \sigma({\rm VBF} \to \chi)}
	{\sigma(gg \to H_{\rm SM}) + \sigma({\rm VBF} \to H_{\rm SM})}
	= \frac{v^2}{f^2} \frac{R_g \sigma(gg \to H_{\rm SM}) + \sigma({\rm VBF} \to H_{\rm SM})}
	{\sigma(gg \to H_{\rm SM}) + \sigma({\rm VBF} \to H_{\rm SM})}.
	\label{eq:xsecratio}
\end{equation}
We take the SM gluon fusion and VBF cross sections from Ref.~\cite{xsecwg}, which includes the current state-of-the-art radiative corrections.\footnote{This approach again allows us to incorporate the full set of currently-known radiative corrections except as described in footnote~\ref{fn:rcs}.}
We then multiply by the appropriate ratio of branching ratios, ${\rm BR}(\chi \to X)/{\rm BR}(H_{\rm SM} \to X)$ for final state $X$, computed as in Sec.~\ref{sec:couplings}.  

\begin{figure}
\includegraphics[scale=0.5]{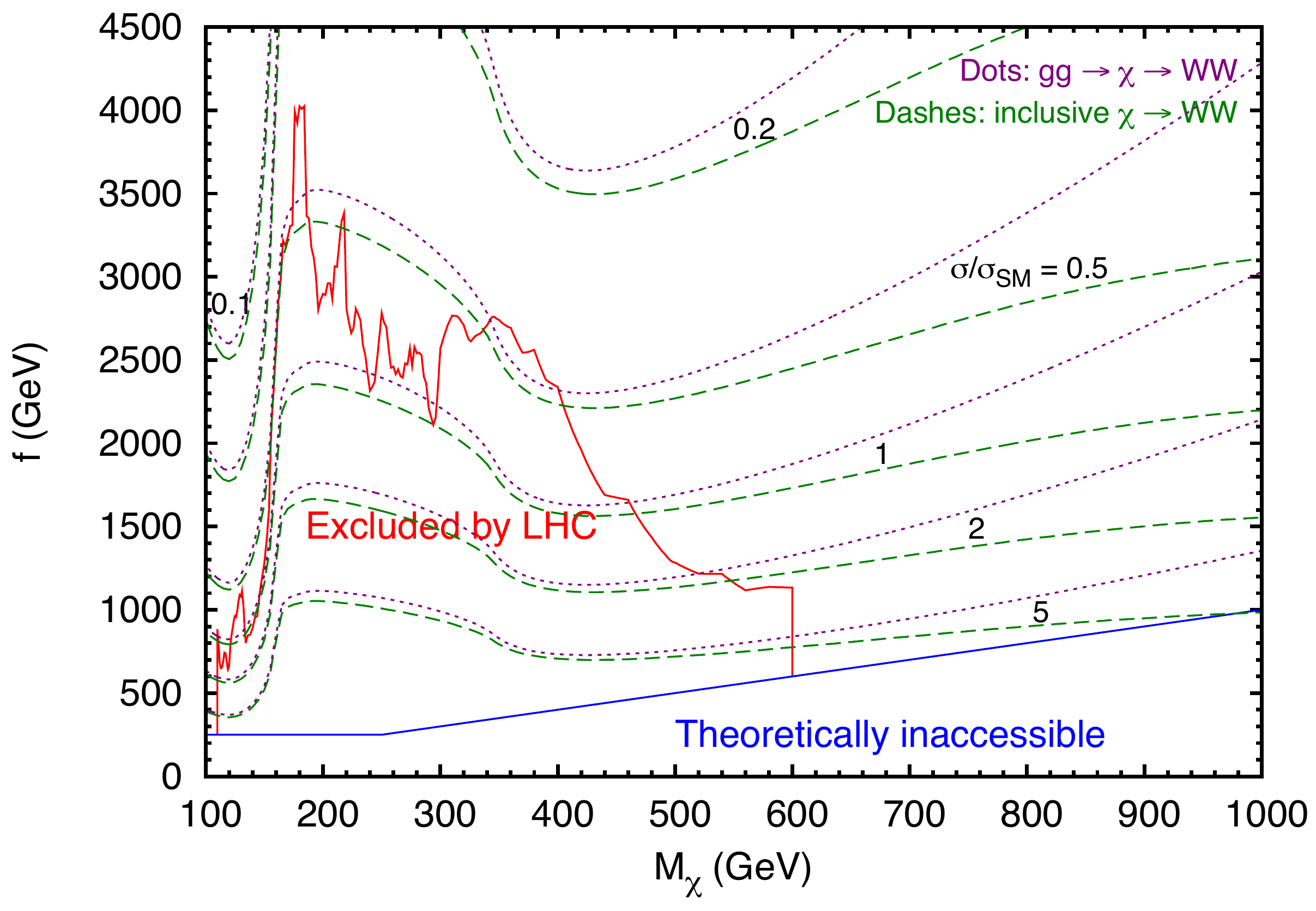}
\caption{Constraints on the dilaton mass and conformal scale $f$ from the LHC.  The area below the solid red line is excluded by the LHC SM Higgs search combined limit or the limit from the $\gamma\gamma$ channel alone; for each mass value we take the stronger limit of ATLAS or CMS.  The region below the solid blue line has $f < M_{\chi}$ or $f < v = 246$~GeV and is thus theoretically unmotivated.  We also show contours of $\sigma({\rm inclusive} \ \chi \to WW)/\sigma({\rm inclusive} \ H_{\rm SM} \to WW)$ (green dashes) and $\sigma(gg \to \chi \to WW)/\sigma(gg \to H_{\rm SM} \to WW)$ (purple dots).}
\label{fig:contours}
\end{figure}

For masses above 150~GeV, the Higgs limits from the LHC rely exclusively on the $WW$ and $ZZ$ channels.  Because ${\rm BR}(\chi \to WW)/{\rm BR}(H_{\rm SM} \to WW) = {\rm BR}(\chi \to ZZ)/{\rm BR}(H_{\rm SM} \to ZZ)$, we can translate the \emph{combined} SM Higgs exclusion into a dilaton exclusion by multiplying the ratio in Eq.~(\ref{eq:xsecratio}) by ${\rm BR}(\chi \to WW)/{\rm BR}(H_{\rm SM} \to WW)$.
Below 150~GeV, the $\gamma\gamma$ channel plays a non-negligible role in the combined Higgs exclusion from both ATLAS and CMS.  However, because ${\rm BR}(\chi \to \gamma\gamma)/{\rm BR}(H_{\rm SM} \to \gamma\gamma) > {\rm BR}(\chi \to WW)/{\rm BR}(H_{\rm SM} \to WW)$, translating the combined SM Higg exclusion into a dilaton exclusion by multiplying the ratio in Eq.~(\ref{eq:xsecratio}) by ${\rm BR}(\chi \to WW)/{\rm BR}(H_{\rm SM} \to WW)$ is \emph{conservative} in this mass range.\footnote{The combined limit also includes small contributions from (i) inclusive Higgs production with decays to $\tau\tau$ below 150 (140)~GeV for ATLAS (CMS), and (ii) associated $WH$ production with decays to $b \bar b$ below 130 (135)~GeV for ATLAS (CMS).  Applying the $WW$ branching ratio scaling to the $\tau\tau$ limit is exact because ${\rm BR}(\chi \to \tau\tau)/{\rm BR}(H_{\rm SM} \to \tau\tau) = {\rm BR}(\chi \to WW)/{\rm BR}(H_{\rm SM} \to WW)$.  The $b\bar b$ contribution, however, comes from a production mode that scales with $v^2/f^2$ and is thus significantly suppressed for the dilaton compared to inclusive production, so that scaling its contribution to the combined exclusion as we have done is not strictly conservative.  Fortunately, the dilaton mass range in which this channel is included in the combined limit will receive a stronger constraint from the $\gamma\gamma$ channel alone.}  
The excluded dilaton parameter space based on the combined SM Higgs exclusion is shown (for masses below 200~GeV) by the solid red line in Fig.~\ref{fig:lep}.
We also compute an exclusion based on the ATLAS and CMS $\gamma\gamma$ channel alone~\cite{ATLASgagaLP11,CMSgagaLP11} by multiplying the ratio in Eq.~(\ref{eq:xsecratio}) by ${\rm BR}(\chi \to \gamma\gamma)/{\rm BR}(H_{\rm SM} \to \gamma\gamma)$.  This is shown by the dot-dashed red line in Fig.~\ref{fig:lep}.  The $\gamma\gamma$ channel provides a stronger exclusion than the (conservative) combined Higgs limit for dilaton masses below 134~GeV.
The excluded dilaton parameter space for masses up to 1000~GeV is shown in Fig.~\ref{fig:contours}.

For dilaton masses above the turn-on of $WW$ decays, the LHC limits are very constraining.  For example, the conformal breaking scale $f$ is constrained to be above 1~TeV (2~TeV) for dilaton masses in the range 145--600~GeV (155--420~GeV).  Below the $WW$ threshold, the LHC constraints are weaker due to the suppression of the detectable dilaton decay branching ratios by the partial width to $gg$.  Nevertheless, $f$ values below 600~GeV are excluded by the current LHC limits, pushing the scale of conformal breaking more than a factor of two above the scale $v$ of EWSB for dilaton masses above 110~GeV.  Below 110~GeV, the limits on $f$ come solely from LEP, yielding $f > 400$~GeV as discussed above.  Combined with the LHC limits, this excludes the possibility that $f \sim v$, as would be expected if the same operators break the conformal and electroweak symmetries.

\section{LHC discovery potential}
\label{sec:discovery}

A dilaton with relatively low conformal breaking scale $f$ is still allowed at low (below $\sim$150~GeV) and high (above $\sim$400~GeV) masses.  We now consider strategies for discovery and characterization of the dilaton properties at the LHC in these two mass ranges.

\subsection{Low-mass dilaton}

The most promising search channel for a low-mass dilaton (below about 135~GeV) is in $\chi \to\gamma\gamma$.  The current LHC and LEP exclusions still allow significant regions of parameter space in which the dilaton $\gamma\gamma$ signature is enhanced relative to the corresponding SM Higgs signal due to the enhanced gluon-fusion production cross section.  Contours of the rate for $gg \to \chi \to \gamma\gamma$ relative to the corresponding SM process are shown by the dotted black lines in Fig.~\ref{fig:lep}.  In particular, because of the weakening of the LEP constraint, a signal in $gg \to \chi \to \gamma\gamma$ as much as 10 times larger than the SM Higgs expectation is still possible for masses between 40 and 110~GeV.  While the $\gamma\gamma$ background increases with decreasing $\gamma\gamma$ invariant mass, the SM Higgs gluon fusion cross section increases as well.  We therefore encourage the LHC experiments to extend the search for a bump in the $\gamma\gamma$ invariant mass spectrum down below 110~GeV to search for a dilaton in this range.  Pushing the LHC limit in the $\gamma\gamma$ channel down to the SM rate would exclude $f$ values up to 1.3~TeV (2.5~TeV) for $M_{\chi} \sim 120$~GeV (40~GeV).

\begin{figure}
\includegraphics[scale=0.5]{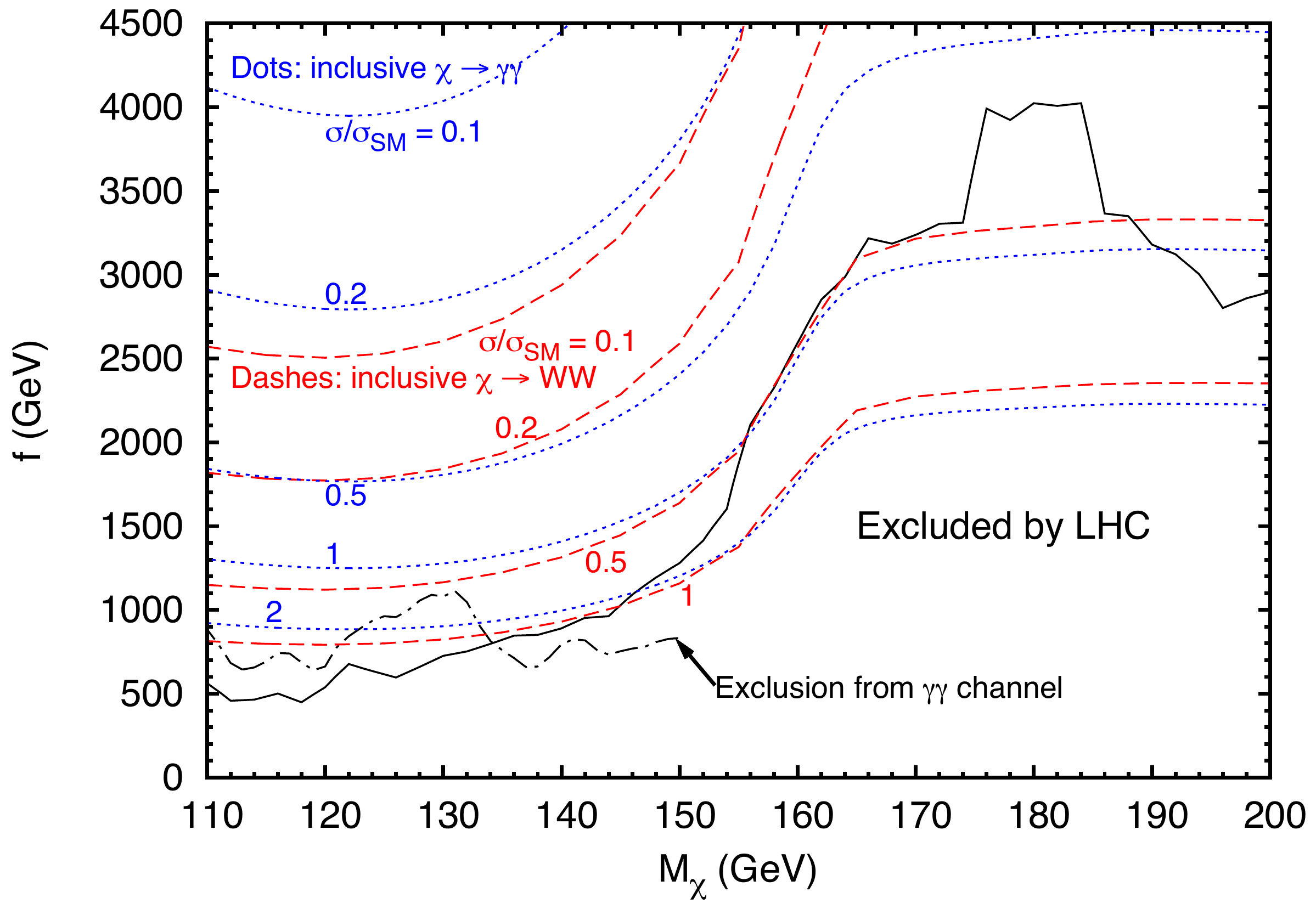}
\caption{Contours of $\sigma({\rm inclusive} \ \chi \to WW)/\sigma({\rm inclusive} \ H_{\rm SM} \to WW)$ (red dashes) and $\sigma({\rm inclusive} \ \chi \to \gamma\gamma)/\sigma({\rm inclusive} \ H_{\rm SM} \to \gamma\gamma)$ (blue dots).  The region below the black curves is excluded by LHC SM Higgs searches.}
\label{fig:contours-low}
\end{figure}

In the event of a discovery in the $\gamma\gamma$ channel, we will want to distinguish the newly-discovered resonance from the SM Higgs and characterize it as a dilaton.  For masses above $\sim$135~GeV, comparing the signal rate in $\gamma\gamma$ to that in $WW$ would immediately indicate a departure from the SM; the relative rates in these two channels differ by more than a factor of two compared to the SM, as shown by the contours in Fig.~\ref{fig:contours-low} (we use Eq.~(\ref{eq:xsecratio}) for the inclusive production cross section scaling).  For lower masses, the suppression of $W\chi$, $Z\chi$, vector boson fusion, or $t \bar t \chi$ channels compared to the SM Higgs expectation would allow the SM hypothesis to be excluded.  For resonance masses below 114.4~GeV, the SM Higgs possibility is of course already excluded by LEP.  

Characterizing a newly-discovered resonance as the dilaton is actually fairly straightforward if it is detectable in a few additional channels.
For $M_{\chi} \gtrsim 135$~GeV, comparison of the signal rates in $\gamma\gamma$ and $WW$ final states provides a measurement of $R_{\gamma}$:
\begin{equation}
	\frac{\sigma({\rm inclusive} \to \chi \to \gamma\gamma)/\sigma({\rm inclusive} \to \chi \to WW)}
	{\sigma({\rm inclusive} \to H_{\rm SM} \to \gamma\gamma)/
		\sigma({\rm inclusive} \to H_{\rm SM} \to WW)}
	= R_{\gamma}.
\end{equation}
This provides a direct test of the QED beta function contribution to the $\chi\gamma\gamma$ coupling via Eq.~(\ref{eq:R}).  An identical measurement can be made for lower dilaton masses by replacing the $WW$ final state with $\tau\tau$.  Measuring $R_g$ requires detection of the dilaton in a production mode other than gluon fusion.  The most promising channels are probably vector boson fusion and associated $W\chi$, $Z\chi$ production, with decays to $\tau\tau$.  This yields,
\begin{equation}
	\frac{\sigma(gg \to \chi \to \tau\tau)/\sigma(W\chi, \chi \to \tau\tau)}
	{\sigma(gg \to H_{\rm SM} \to \tau\tau)/\sigma(WH_{\rm SM}, H_{\rm SM} \to \tau\tau)}
	= R_g,
\end{equation}
and similarly for the $Z\chi$ and vector boson fusion production modes (these modes could be combined for greater statistical power).  As before, this provides a direct test of the QCD beta function contribution to the $\chi gg$ coupling via Eq.~(\ref{eq:R}).  The universality of the scaling of the dilaton couplings to all SM particles other than $gg$ and $\gamma\gamma$ can be checked by verifying that
\begin{eqnarray}
	\frac{\sigma(X \to \chi \to Y)}{\sigma(X \to \chi \to Z)} 
	&=& \frac{\sigma(X \to H_{\rm SM} \to Y)}{\sigma(X \to H_{\rm SM} \to Z)} \qquad 
	{\rm for} \ Y,Z \neq \gamma\gamma, \ {\rm and} \nonumber \\
	\frac{\sigma(X \to \chi \to Z)}{\sigma(Y \to \chi \to Z)}
	&=& \frac{\sigma(X \to H_{\rm SM} \to Z)}{\sigma(Y \to H_{\rm SM} \to Z)} \qquad
	{\rm for} \ X,Y \neq \ {\rm gluon \ fusion}.
\end{eqnarray}

The measurement of $R_g$ will be very challenging because the dilaton production cross sections in the $W\chi$, $Z\chi$, and vector boson fusion channels are suppressed by $v^2/f^2$ compared to the corresponding SM Higgs cross sections; combining this with the suppression of the visible decay branching ratios and existing lower bounds on $f$ yields suppressions of these channels by at least a factor of 20 compared to the SM Higgs prediction.  An upper bound on the dilaton rate in the $W\chi$, $Z\chi$, and vector boson fusion channels would set a lower bound on $R_g$.

The measurements of $R_{\gamma}$ and $R_g$ alone do not allow one to calculate the new particle's branching ratios in a model independent way; however, they should allow enough confidence in the dilaton nature of the new particle for the branching ratios to be computed under this assumption.  With this theory assumption, the conformal breaking scale $f$ can be obtained using
\begin{equation}
	\frac{\sigma(gg \to \chi \to \gamma\gamma)}{\sigma(gg \to H_{\rm SM} \to \gamma\gamma)} 
	= R_g \frac{v^2}{f^2} \times
	\frac{{\rm BR}(\chi \to \gamma\gamma)}{{\rm BR}(H_{\rm SM} \to \gamma\gamma)},
\end{equation}
or from the corresponding inclusive $\gamma\gamma$ cross sections by replacing $R_g v^2/f^2$ above with the right-hand side of Eq.~(\ref{eq:xsecratio}).

\subsection{High-mass dilaton}

\begin{figure}
\includegraphics[scale=0.5]{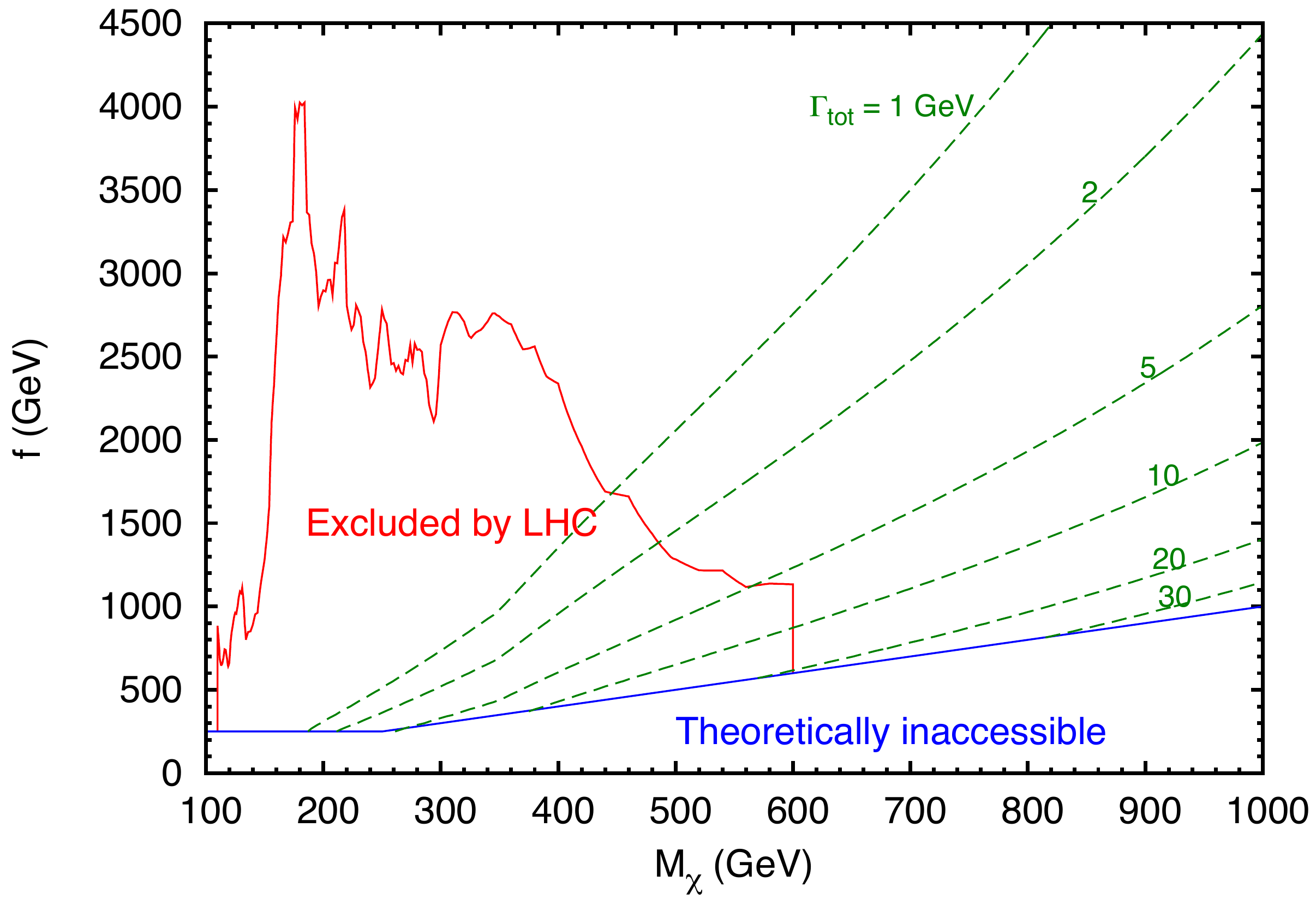}
\caption{Contours of the dilaton total width.  The region below the solid red line is excluded by LHC SM Higgs searches, while the region below the solid blue line has $f < M_{\chi}$ or $f < 246$~GeV and is thus theoretically unmotivated.  Note that the total dilaton width is less than 40~GeV in the entire allowed parameter space shown.}
\label{fig:widthcontours}
\end{figure}

For a high-mass dilaton, the most sensitive search channels will involve decays to $ZZ$, just as for the SM Higgs.  The dilaton, however, has a much narrower total width in this mass range than the SM Higgs, as illustrated in Fig.~\ref{fig:widthcontours}.  This will make the dilaton much easier to discover in the $ZZ \to 4\ell$ channel because the signal will form a narrow peak, allowing backgrounds to be reduced through a tighter invariant mass cut than in the SM Higgs case.  It will also make the dilaton easier to discover in that the theoretical uncertainty due to interference between the signal and SM backgrounds is dramatically reduced.  This theoretical uncertainty due to the large SM Higgs width was parameterized as $150 \times (M_H \ {\rm [TeV]})^3$\% for $M_H \geq 300$~GeV\footnote{For comparison, a 300~GeV SM Higgs boson has a width of about 8.5~GeV~\cite{HDECAY}.  The SM Higgs width grows with $M_H^3$ in the high mass range.} in the ATLAS $H \to ZZ \to \ell\ell\nu\nu$ analysis of Ref.~\cite{ATLAS-CONF-2011-148}.  Because the dilaton width is very closely equal to $v^2/f^2$ times the SM Higgs width, this theoretical uncertainty could be replaced in the dilaton case with $150 \times (M_{\chi} \ {\rm [TeV]})^3 \times v^2/f^2$\%.  

To estimate the LHC discovery potential for a high-mass, narrow-width dilaton in the $ZZ \to 4\ell$ channel, we studied the $pp\rightarrow\chi\rightarrow ZZ\rightarrow 4l$ signal and corresponding background using {\tt CALCHEP}~\cite{calchep}.  We added the dilaton couplings to {\tt CALCHEP} including the leading-order $gg \to \chi$ effective coupling [proportional to the numerator of $R_g$ in Eq.~(\ref{eq:R})].  We generated 100,000 events in the $e^+e^-\mu^+\mu^-$ channel for the SM background\footnote{The background to this process was calculated by generating all diagrams contributing to a 4$\ell$ final state \emph{except} the Higgs exchange ones. Also note that since the final state is 4$\ell$, we do not include vector boson fusion diagrams, as these come associated with two additional hard jets.} and 3,000 events for the signal at each point in a grid of $M_{\chi}, f$ for the 7~TeV LHC.  We mimicked detector resolution by smearing the lepton energies according to $\Delta E/E =a \oplus b/\sqrt{E}$, where $a$ and $b$ are $5.5\times10^{-3}$ and $5\times10^{-2}$ for the electrons and $1.3\times10^{-2}$ and $1.5\times10^{-4}$ for the muons. We do not include any $k$-factors.

\begin{figure}
\includegraphics[scale=0.8]{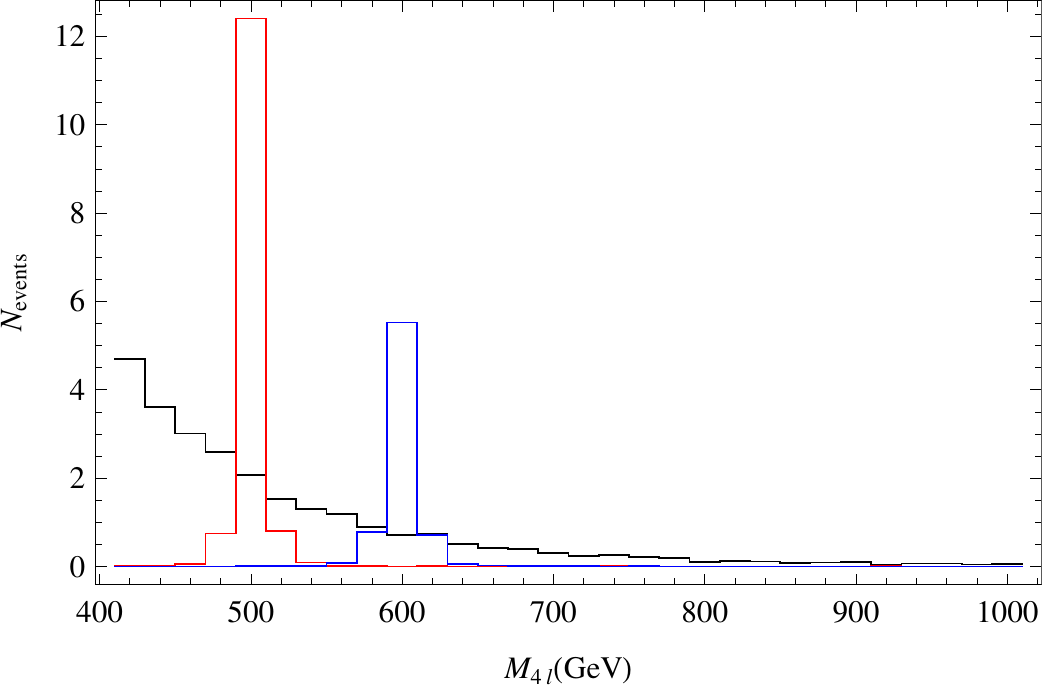}
\includegraphics[scale=0.8]{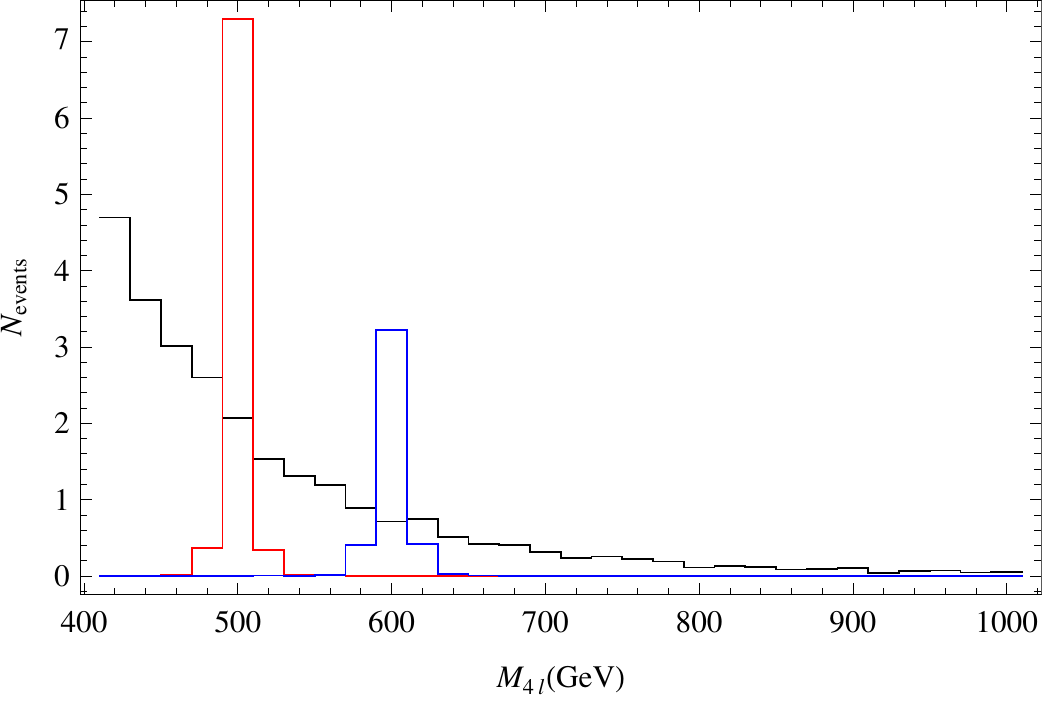}
\caption{Distribution of the reconstructed $4\ell$ invariant mass after cuts for the SM background and dilaton signal, for dilaton masses of 500 and 600~GeV and $f$ values of 1.5~TeV (left) and 2~TeV (right).  Event numbers are shown for a luminosity of 50~fb$^{-1}$ at the 7~TeV LHC; the bin width is 20~GeV.}
\label{fig:dist}
\end{figure}

We apply the following selection cuts:
\begin{itemize}
\item $p_T > 10$~GeV and $|\eta| < 2.5$ for each of the four leptons;
\item $80~{\rm GeV} < M_{ee}, M_{\mu\mu} < 100~{\rm GeV}$ (this reduces the $ee\mu\mu$ background not coming from two on-shell $Z$ bosons).
\end{itemize}
The resulting $4\ell$ invariant mass distributions for signal and background are shown for a few dilaton masses and $f$ values in Fig.~\ref{fig:dist}.  We have incorporated the $4e$ and $4\mu$ final states by multiplying the simulated signal and background cross sections by two.  The dilaton signal peaks are quite narrow and the cross section is appreciable.  Compared to the much wider SM Higgs, the narrowness of the dilaton offers the additional benefit of being able to measure the background from data using sidebands.

\begin{figure}
\includegraphics[scale=0.8]{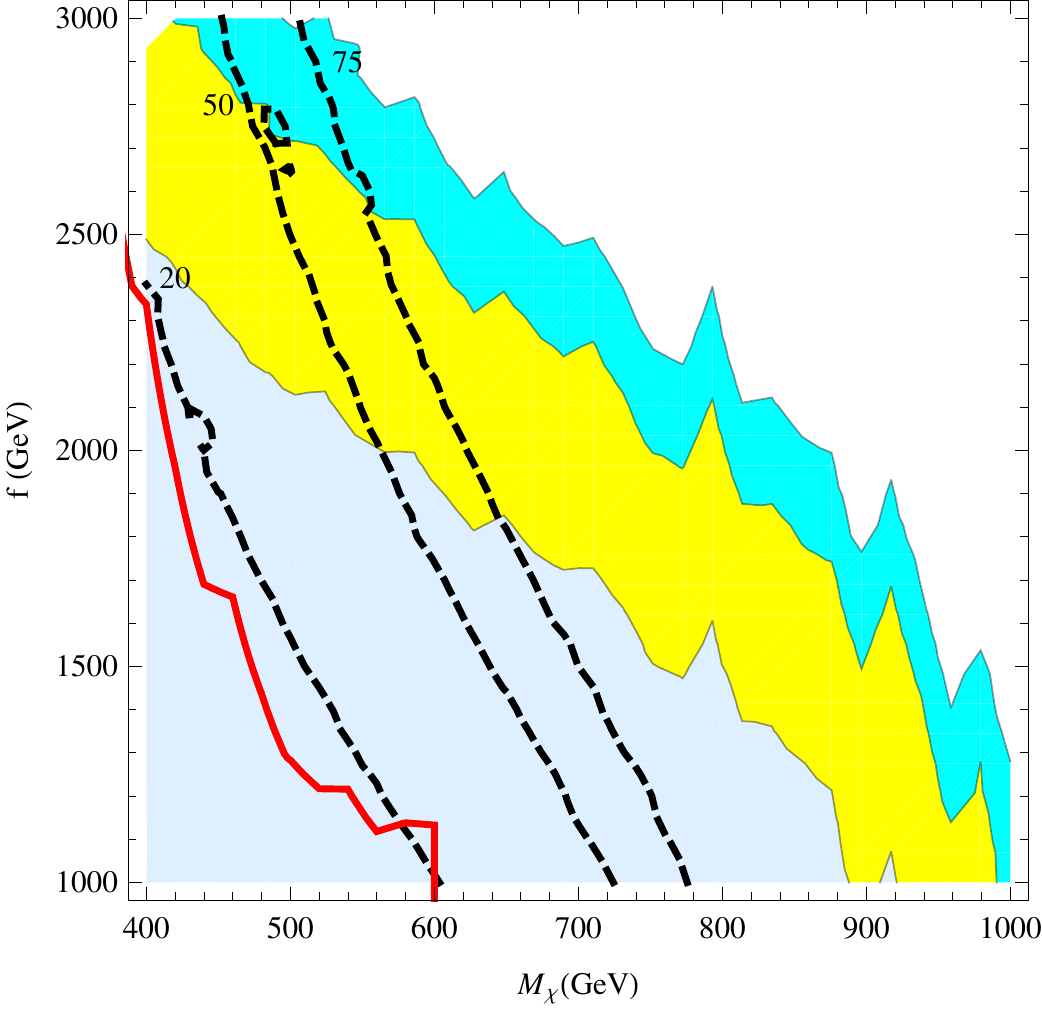}
\hspace{0.05in}
\includegraphics[scale=0.8]{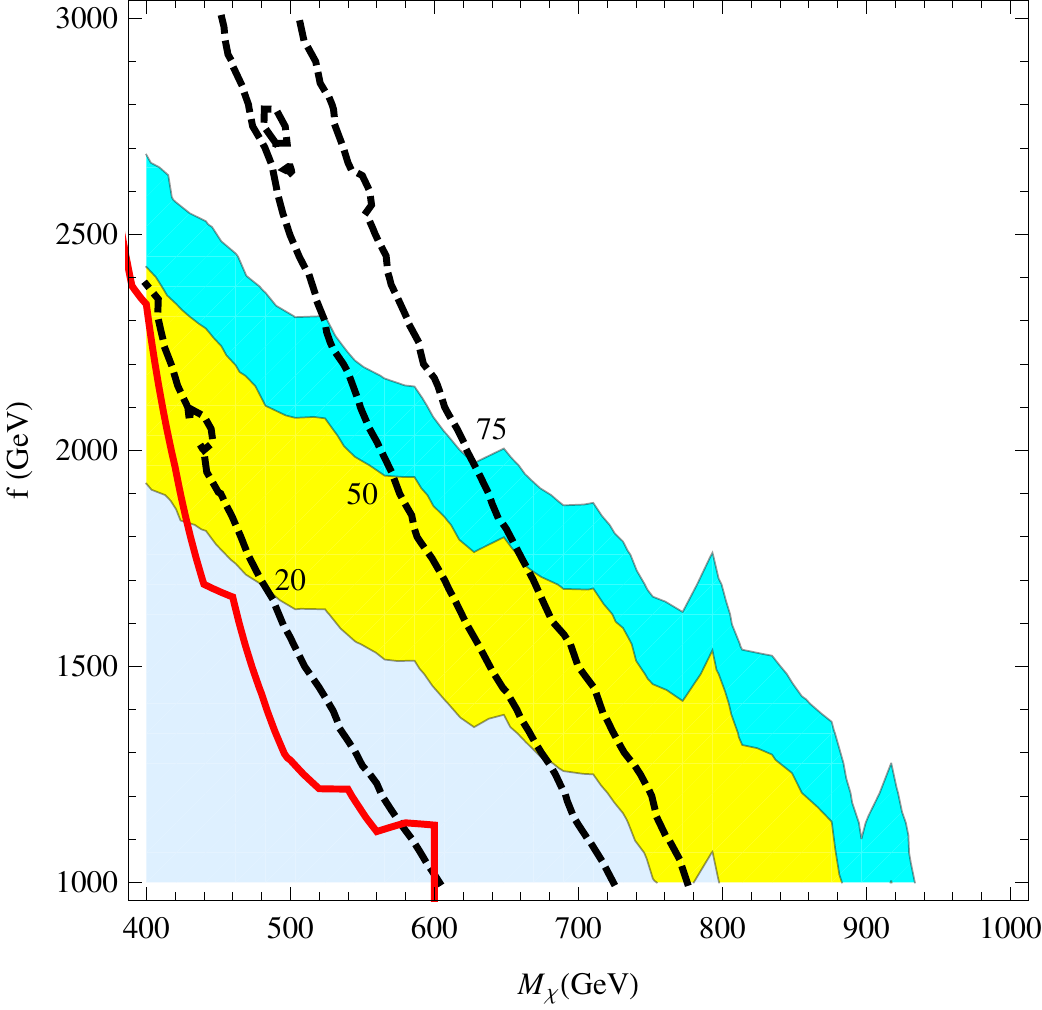}
\caption{Luminosity required at the 7~TeV LHC for observing the dilaton in the $pp\rightarrow4\ell$ process as a 3$\sigma$ evidence (left) or 5$\sigma$ discovery (right). The thick dashed lines are contours of 5 signal events for luminosities (from left to right) of 20, 50, and 75~fb$^{-1}$. It is only the regions below these dashed lines that are truly observable at the LHC for the corresponding luminosities---the regions above them do not have enough signal events. Regions to the left of the solid red line are already excluded by the LHC Higgs search.}
\label{fig:lum1}
\end{figure}

\begin{figure}[h!]
\includegraphics[scale=0.8]{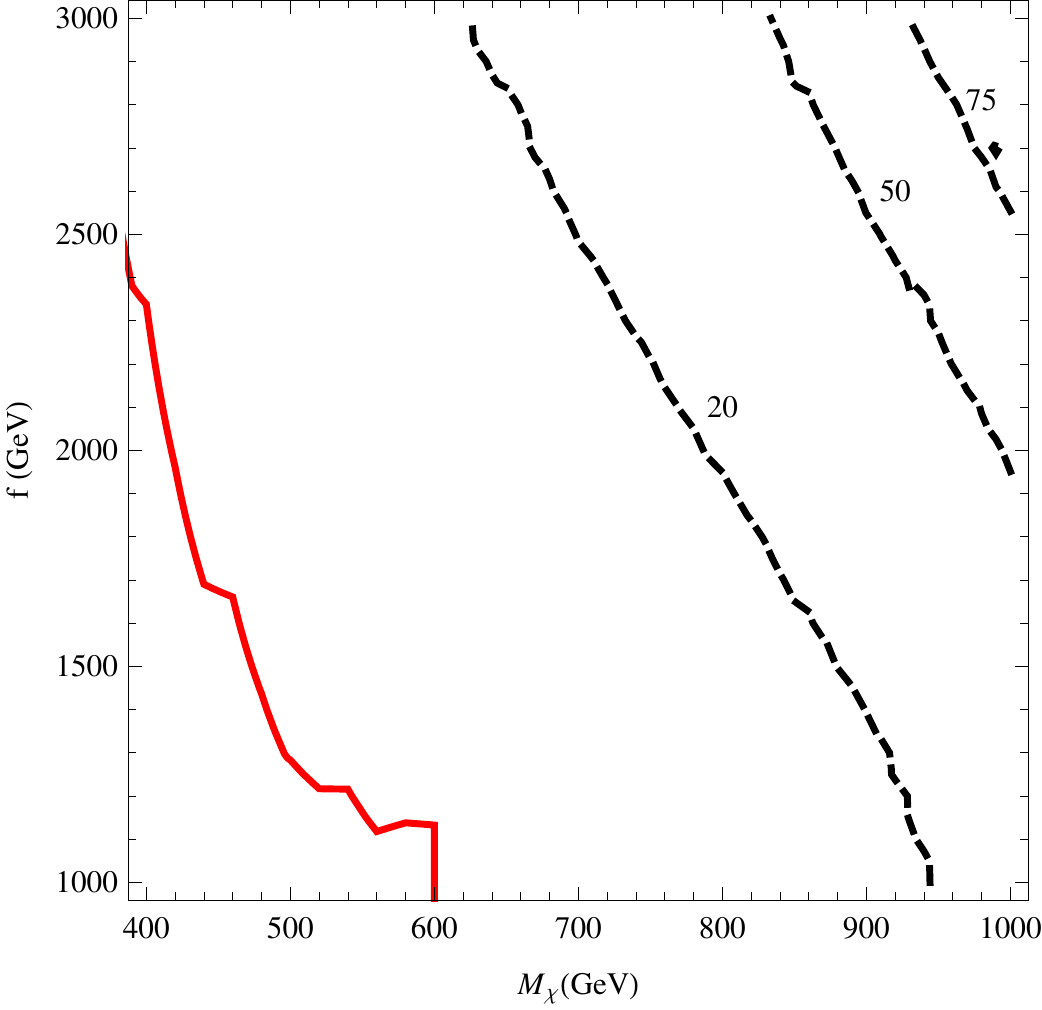}
\caption{Luminosity required for 5 signal events for dilaton discovery in the $ZZ$ channel at the 14~TeV LHC. The region to the left of each contour is discoverable with the specified luminosity in fb$^{-1}$.  Regions to the left of the solid red line are already excluded by the LHC Higgs search.}
\label{fig:lum2}
\end{figure}

To determine the discovery sensitivity, we finally apply a tight cut on the four-lepton invariant mass,
\begin{itemize}
\item $M_{\chi} - 10~{\rm GeV} < M_{4\ell} < M_{\chi} + 10~{\rm GeV}$.
\end{itemize}
The resulting luminosity required for a 3$\sigma$ or 5$\sigma$ dilaton discovery sensitivity at the 7~TeV LHC is shown in Fig.~\ref{fig:lum1}.  The shaded regions show contours of the luminosity required for $N_S/\sqrt{N_B} \geq 3$ (left plot) or 5 (right plot), corresponding to a Gaussian statistical sensitivity of 3$\sigma$ or 5$\sigma$, where $N_S$ and $N_B$ are the numbers of signal and background events that survive the cuts.  For an observation or discovery we also require at least 5 signal events; contours of $N_S = 5$ are superimposed on Fig.~\ref{fig:lum1} as thick dashed lines---only the region of parameter space below the corresponding dashed line should be considered observable for each luminosity.  Dilaton masses up to 640~GeV should be discoverable at the 7 TeV LHC with 50~fb$^{-1}$ for $f = 1.5$~TeV. For the 14~TeV LHC, the entire parameter space considered has $N_S/\sqrt{N_B} \geq 5$ for $\mathcal{L} \geq 20$~fb$^{-1}$, and the discovery sensitivity is only limited by the availability of enough signal events. We show the 5-event luminosity contours for the 14~TeV machine in Fig.~\ref{fig:lum2}---the region to the left of the contour is discoverable for each particular luminosity.

We finally consider how to characterize a newly-discovered resonance as the dilaton in the high mass range.  First, the new state is easily distinguishable from a SM Higgs in this mass range by its significantly narrower width and the suppression of the vector boson fusion production mode.  If the total width of the dilaton is large enough to be resolved experimentally in the $4\ell$ lineshape, the conformal breaking scale $f$ can be determined directly using\footnote{Electroweak radiative corrections to the SM Higgs total width become significant in the high mass range; care must be taken to separate out any corrections due to the large triple-Higgs coupling.}
\begin{equation}
	\frac{\Gamma_{\rm tot}(\chi)}{\Gamma_{\rm tot}(H_{\rm SM})} \simeq \frac{v^2}{f^2},
\end{equation}
which holds to better than 5\% for $M_{\chi} > 300$~GeV.
If the rate for dilaton production via vector boson fusion is large enough to be detected, $f$ can also be obtained from
\begin{equation}
	\frac{\sigma({\rm VBF} \to \chi \to ZZ)}{\sigma({\rm VBF} \to H_{\rm SM} \to ZZ)}
	\simeq \frac{v^2}{f^2},
\end{equation}
where we have used the assumption that ${\rm BR}(\chi \to ZZ)$ is the same as that of the SM Higgs, which again holds to better than 5\% for $M_{\chi} > 300$~GeV.  Finally, if the rate for dilaton production via vector boson fusion is large enough to be detected, $R_g$ can be measured using
\begin{equation}
	\frac{\sigma(gg \to \chi \to ZZ)/\sigma({\rm VBF} \to \chi \to ZZ)}
	{\sigma(gg \to H_{\rm SM} \to ZZ)/\sigma({\rm VBF} \to H_{\rm SM} \to ZZ)}
	= R_g.
\end{equation}
As in the case of a light dilaton, this provides a direct test of the QCD beta function contribution to the $\chi gg$ coupling via Eq.~(\ref{eq:R}).

\section{ILC prospects}
\label{sec:ilc}

The ILC~\cite{Djouadi:2007ik} offers excellent prospects for measuring the couplings and other properties of a light SM-like Higgs boson.  These can be applied to the dilaton as follows.  The total production cross section for $e^+e^- \to Z \chi$ can be measured independently of the $\chi$ decay modes using the recoil mass technique~\cite{GarciaAbia:1999kv}.  This directly determines $v^2/f^2$.  The dilaton branching ratios can then be measured in a model-independent way using the event sample selected by the recoil mass.  Hadronic decays can be separated into $b\bar b$, $c \bar c$, and $gg$ samples using $b$- and charm-tagging.  For $M_{\chi} < 160$~GeV, after determining that the largest decay branching ratio is to $gg$, a measurement of ${\rm BR}(\chi \to b \bar b)$ and/or ${\rm BR}(\chi \to WW)$ will allow $R_g$ to be extracted.\footnote{The measurement of a subdominant branching ratio is necessary to measure $R_g$; a lower bound on ${\rm BR}(\chi \to gg)$ would allow only a lower bound on $R_g$ to be set.  The larger of ${\rm BR}(\chi \to b \bar b)$ and ${\rm BR}(\chi \to WW)$ is at least 4\% over the whole mass range being considered.}

\begin{figure}
\includegraphics[scale=0.35]{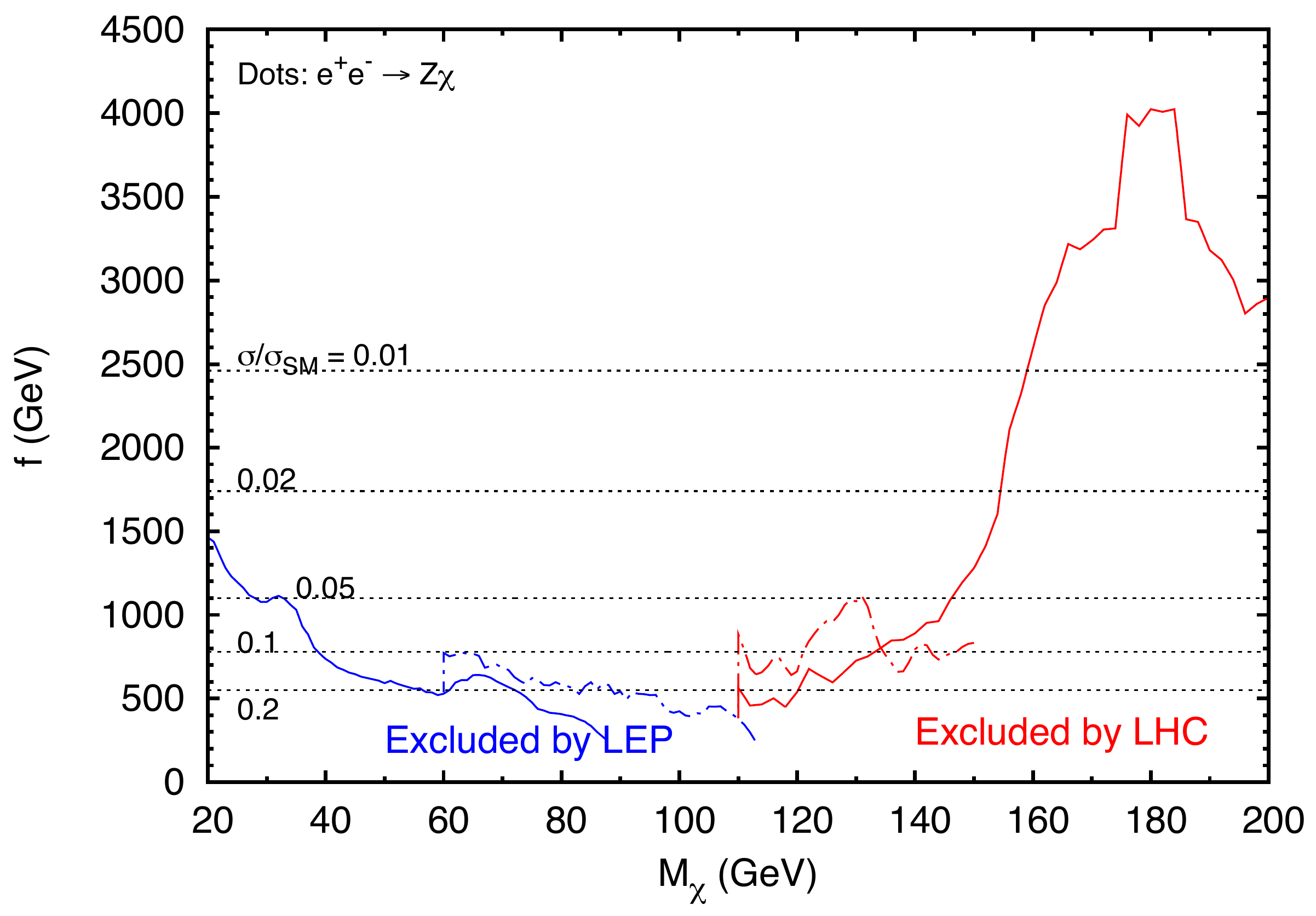}
\includegraphics[scale=0.35]{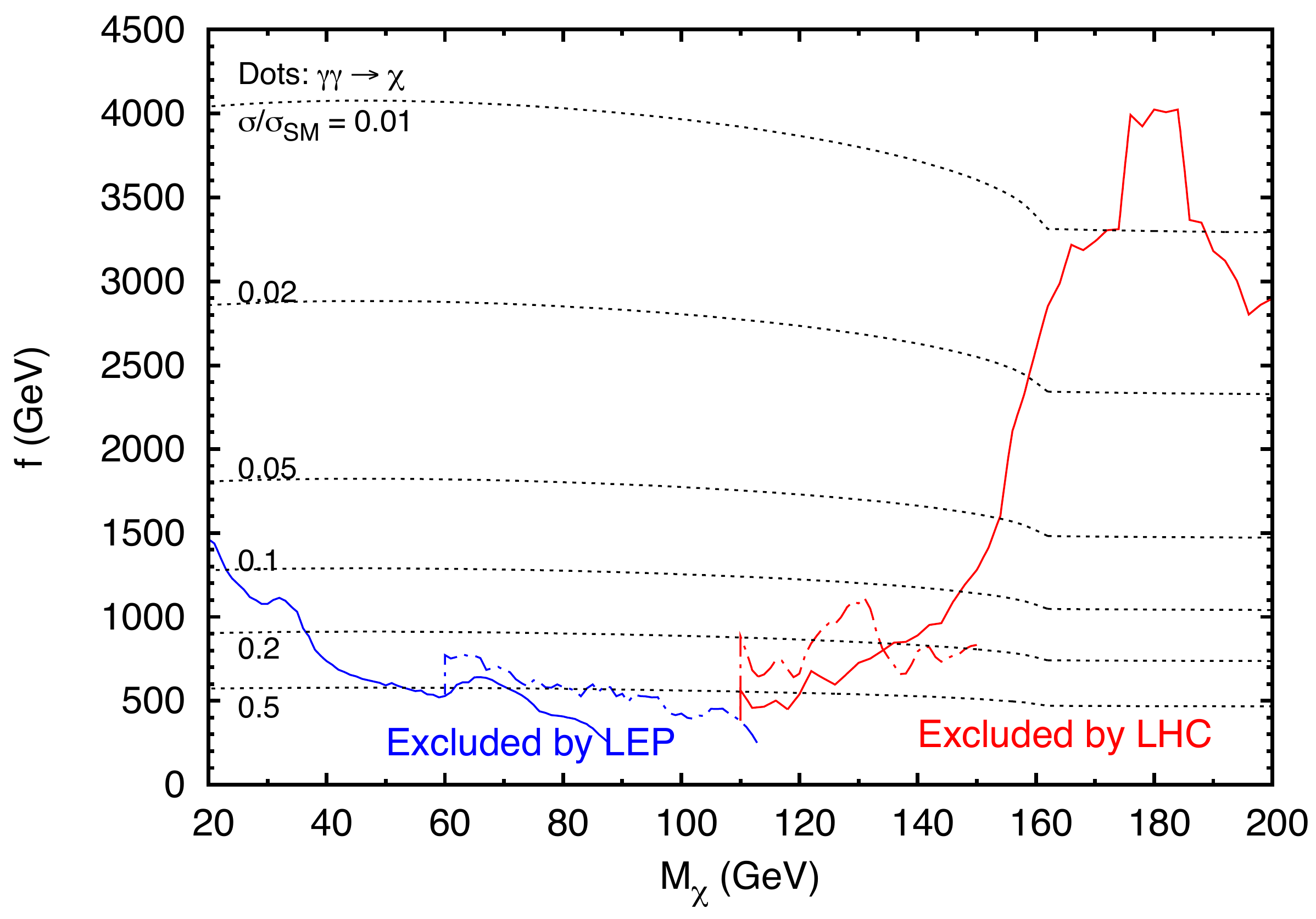}
\caption{Dilaton production cross sections relative to those of the SM Higgs at an $e^+e^-$ collider (left) and photon collider (right).  The regions below the blue and red lines are excluded as in Fig.~\ref{fig:lep}.}
\label{fig:ilc}
\end{figure}

Unfortunately, a major difficulty with this approach is that the cross section for $e^+e^- \to Z\chi$ is suppressed by a factor of $v^2/f^2$ compared to the corresponding SM Higgs cross section.  The other ILC dilaton production modes, $e^+e^- \to \bar \nu \chi \nu$ (via $WW$ fusion), $e^+e^- \to e^+ \chi e^-$ (via $ZZ$ fusion), and even $e^+e^- \to t \bar t \chi$ at higher $e^+e^-$ collision energies, are suppressed by the same factor.  Based on the current LEP and LHC exclusions, ILC production of a dilaton lighter than 120~GeV would be suppressed by a factor of at least 5--10, while production of a dilaton between 120 and 145~GeV would be suppressed by a factor of at least 10--20 (see the left panel of Fig.~\ref{fig:ilc}), severely reducing the signal statistics available for cross section and branching ratio measurements.

The cross section situation is better at a photon collider because of the relative enhancement factor $R_{\gamma}$ in the $\gamma\gamma \to \chi$ cross section; nevertheless, production rates larger than about half the corresponding SM Higgs rate are already excluded (see the right panel of Fig.~\ref{fig:ilc}).  The most interesting channels are $b \bar b$ (for lower masses)~\cite{Niezurawski:2003iu} and $WW,ZZ$ (for higher masses)~\cite{Niezurawski:2004ui}.  The significant suppression of these final states by the $gg$ decay below $2M_W$ makes the situation even more difficult.  Detection of the $gg$ final state itself is probably unfeasible at a photon collider due to the large $q \bar q$ background.

\section{Conclusions}
\label{sec:conclusions}
The LHC should soon give us clues about the physics responsible for electroweak symmetry breaking. Among the various possibilities, EWSB through a strongly coupled conformal dynamics is an interesting avenue from various perspectives. If this conformal dynamics is broken spontaneously, a dilaton, the Goldstone boson associated with the spontaneous breaking of scale invariance, could emerge in the low-energy spectrum. Such a state is very interesting to study from two points of view. First, its couplings are  similar to those of the Higgs (indeed the Standard Model Higgs itself can be thought of as a dilaton~\cite{Goldberger:2007zk}), so a situation could arise in which a light scalar with properties similar to a Higgs is discovered, while it is not in fact responsible for electroweak symmetry breaking. Secondly, the dilaton if identified properly could provide a hint to the conformal nature of the strong sector which might otherwise be difficult to establish.

In this paper, we recast the LEP and LHC exclusion limits for a Standard Model Higgs into exclusion regions in the two-dimensional $M_\chi-f$ dilaton parameter space. We find that for low values of $f$, the dilaton is already excluded by the LHC in a large portion of parameter space. This is due to an enhancement in the coupling of the dilaton to two gluons relative to the SM Higgs, which will always be present if QCD is itself part of the conformal dynamics at high energies. For large $f$ and large $M_\chi$, the dilaton is not excluded but could be discovered at LHC with more luminosity. This dilaton would in fact be easier to find than a Standard Model Higgs because it is much more narrow, its width being suppressed by $(v/f)^2$. A low-mass dilaton is still allowed for relatively small values of $f$, and could still yield a $\gamma \gamma$ signature that is larger than that of the SM Higgs. In fact, because the dilaton has a suppressed cross-section in the $Z \chi$ mode, the LEP bound on the Standard Model Higgs does not apply to the dilaton. Therefore a very light dilaton is still allowed, and could be discovered  at the LHC in the $\gamma \gamma$ channel if the Standard Model Higgs search is extended below the LEP Higgs mass bound. 
\linebreak

\textbf {Note added}: While we were finishing this paper, Ref.~\cite{Barger:2011nu} appeared which has significant overlap with our results.

\begin{acknowledgments}
This work was supported by the Natural Sciences and Engineering Research Council of Canada. 
\end{acknowledgments}


\end{document}